\documentclass[journal]{IEEEtran}
\usepackage{xcolor,soul,framed} 

\colorlet{shadecolor}{yellow}
\usepackage[pdftex]{graphicx}
\graphicspath{{../pdf/}{../jpeg/}}
\DeclareGraphicsExtensions{.pdf,.jpeg,.png}

\usepackage[cmex10]{amsmath}
\usepackage{cite}
\usepackage{booktabs}

\usepackage{flushend}
\usepackage{rotating}
\usepackage{hyperref} 
\usepackage{tablefootnote}
\usepackage{array}
\usepackage{mdwmath}
\usepackage{amsmath}
\usepackage{amsfonts}
\usepackage{amssymb}
\usepackage{mdwtab}
\usepackage{eqparbox}
\usepackage{url}
\usepackage{xcolor}
\usepackage{comment}
\usepackage{footnote}
\usepackage{multirow}
\usepackage{graphicx}
\usepackage{caption}
\usepackage{algorithm}
\usepackage{algorithmic}
\allowdisplaybreaks

\usepackage{float}
\usepackage{placeins}

\usepackage{url} 
\usepackage{array, makecell}

\usepackage{tikz}

\usetikzlibrary{fit,positioning}
\usetikzlibrary{bayesnet}
\usetikzlibrary{arrows}

\begin{document}
\bstctlcite{IEEEexample:BSTcontrol}
    \title{GLUSE: Enhanced Channel-Wise Adaptive Gated Linear Units SE for Onboard Satellite Earth Observation Image Classification}
    
     \author{\IEEEauthorblockN{Thanh-Dung Le,~\IEEEmembership{Senior Member,~IEEE}, Vu Nguyen Ha,~\IEEEmembership{Senior Member,~IEEE}, Ti Ti Nguyen,~\IEEEmembership{Member,~IEEE}, Duc-Dung Tran,~\IEEEmembership{Member,~IEEE}, Hung Nguyen-Kha, Luis M. Garces-Socarras,~\IEEEmembership{Member,~IEEE}, 
    Juan Carlos Merlano-Duncan,~\IEEEmembership{Senior Member,~IEEE}, 
    Symeon Chatzinotas,~\IEEEmembership{Fellow,~IEEE} 
    \vspace{-3mm}
    }

    \thanks{This work was funded by the Luxembourg National Research Fund (FNR), with the granted SENTRY project corresponding to grant reference C23/IS/18073708/SENTRY.}

    \thanks{Thanh-Dung Le, Vu Nguyen Ha, Ti Ti Nguyen, Duc-Dung Tran, Hung Nguyen-Kha, Luis M. Garces-Socarras, Juan Carlos Merlano-Duncan, Symeon Chatzinotas are with the Interdisciplinary Centre for Security, Reliability, and Trust (SnT), University of Luxembourg, Luxembourg (Corresponding author. Email: thanh-dung.le@tamucc.edu).} 

    \thanks{This paper is a revised and expanded version of a paper entitled “Semantic Knowledge Distillation for Onboard Satellite Earth Observation Image Classification”, which was presented at IEEE ICMLCN 2025, Barcelona, Spain, 26–29 May 2025.}
}  

\markboth{Accepted for Publication in IEEE Journal of Selected Topics in Applied Earth Observations and Remote Sensing, 2026.
}{Thanh-Dung Le \MakeLowercase{\textit{et al.}}: }

 \maketitle

\begin{abstract}
This study introduces ResNet-GLUSE, a lightweight ResNet variant enhanced with Gated Linear Unit-enhanced Squeeze-and-Excitation (GLUSE), an adaptive channel-wise attention mechanism. By integrating dynamic gating into the traditional SE framework, GLUSE improves feature recalibration while maintaining computational efficiency. Experiments on EuroSAT and PatternNet datasets confirm its effectiveness, achieving exceeding \textbf{94\% and 98\% accuracy}, respectively. While \textbf{MobileViT achieves 99\% accuracy}, ResNet-GLUSE offers \textbf{33× fewer parameters, 27× fewer FLOPs, 33× smaller model size (MB), $\approx$6× lower power consumption (W), and $\approx$3× faster inference time (s)}, making it significantly more efficient for onboard satellite deployment. Furthermore, due to its simplicity, ResNet-GLUSE can be easily mimicked for \textbf{neuromorphic computing}, enabling ultra-low power inference at just \textbf{852.30 mW} on Akida Brainchip. This balance between high accuracy and ultra-low resource consumption establishes ResNet-GLUSE as a practical solution for real-time Earth Observation (EO) tasks. Reproducible codes are available in our shared repository.

\vspace{3pt}
\textbf{\textit{Impact Statement--}} ResNet‑GLUSE adds a lightweight, dynamic gated attention to a small ResNet, reaching high accuracy with a fraction of the computation, memory, and power required by existing models, enabling real‑time image analytics directly on satellite edge hardware. This synergy of performance, scalability, and energy efficiency accelerates rapid decision-making in resource-constrained orbital environments, aiding critical tasks like near-real-time hazard detection and precision agriculture. Reproducible codes promote broad adoption and facilitate ongoing innovation, underscoring ResNet-GLUSE’s potential as a transformative solution for next-generation EO missions.

\end{abstract}

\begin{IEEEkeywords}
Earth Observation, Remote Sensing, Knowledge Distillation, Onboard Processing, Artificial Intelligence, ResNet.
\end{IEEEkeywords}

\IEEEpeerreviewmaketitle


\section{Introduction}

\IEEEPARstart{T}{he} rapid increase in satellite deployments for EO and remote sensing (RS) missions reflects a growing demand for applications like environmental monitoring, disaster response, precision agriculture, and scientific research \cite{sadek2021new}. These applications rely on high-frequency, high-resolution data for timely and accurate decision-making. However, a significant bottleneck in Low Earth Orbit (LEO) satellite operations is reliance on ground stations for data transmission, which limits the availability of communication windows and results in frequent connectivity loss \cite{al2022survey}. This delay can impede critical responses in situations requiring immediate data access.

The advent of Satellite Internet Providers, such as Starlink and OneWeb, offers the potential for continuous (24/7) connectivity to LEO satellites, facilitating on-demand data access \cite{chougrani2024connecting}. Yet, seamless connectivity alone does not fully meet modern EO and RS requirements, which increasingly demand real-time, onboard decision-making. For optimal operations, onboard neural networks (NNs) must prioritize computational efficiency to autonomously analyze data, identify critical information, and make immediate adjustments, such as refocusing on a target area during subsequent satellite passes \cite{fontanesi2023artificial}.

Historically, onboard NNs have been designed for efficiency, often relying on convolutional neural network (CNN) models to balance performance and resource constraints. For example, the $\Phi$-Sat-1 mission used a CNN-based NN for onboard image segmentation using the Intel Movidius Myriad 2 vision processing unit (VPU), representing the first deployment of deep learning on a satellite \cite{giuffrida2021varphi}. Similarly, $\Phi$-Sat-2 adopted a convolutional autoencoder for image compression to reduce transmission requirements, demonstrating the feasibility of lightweight models on hardware-constrained environments on three different hardware, including graphic processing unit (GPU) NVIDIA GeForce GTX 1650, VPU Myriad 2, and central processing unit (CPU) Intel Core i7-6700\cite{guerrisi2023artificial}. 

Despite their efficiency, CNNs can be limited in performance, especially compared to the recent success of Vision Transformer (ViT) architectures. ViTs have gained popularity in computer vision due to their ability to capture global context via self-attention mechanisms, often surpassing traditional CNNs in performance. However, ViTs require significantly more computational power and memory as image resolution increases, which poses challenges for deployment on power-constrained satellite platforms \cite{chou2024semantic, le2024board}. In line with this limitation, NASA has emphasized that the benefits of fine-tuning smaller, lightweight models for oriented tasks currently outweigh the costs and risks associated with large models \cite{2024CN000258}.

ResNet architectures have emerged as a compelling solution by effectively addressing CNNs' vanishing gradient issues through skip connections, achieving a favorable balance between computational efficiency and performance \cite{mascarenhas2021comparison, goldblum2024battle, haruna2025exploring}. Recent studies further illustrate that employing knowledge distillation (KD) \cite{hinton2015distilling} from pretrained ViT models can significantly enhance lightweight ResNet models by transferring semantic knowledge, thereby improving their performance to levels comparable to ViTs, while maintaining practicality for onboard satellite deployment \cite{le2024semantic}.

Motivated by these findings, this study aims to further enhance the lightweight ResNet model through improved channel-wise feature recalibration. Specifically, we propose GLUSE, an adaptive channel-wise attention mechanism inspired by Gated Linear Units (GLU) \cite{shazeer2020glu} integrated into the Squeeze-and-Excitation (SE)  framework \cite{hu2018squeeze}. GLUSE is designed to optimize the performance-complexity trade-off, maintaining suitability for on-the-air deployment. While onboard EO processing encompasses a broad range of vision tasks, including object detection and semantic segmentation\cite{ hoeser2020object, zhuang2026frequency} this study deliberately targets classification (Section \ref{subsec:classification}) and, image reconstruction under realistic satellite transmission conditions (Section \ref{subsec:reconstruction}) as two representative, resource-constrained onboard tasks, rather than positioning classification as a self-sufficient solution to onboard EO processing; detection and segmentation are discussed as important directions for future work in Section \ref{sec:limitations}.

Experimental validations demonstrate that the proposed ResNet-GLUSE consistently surpasses traditional SE methods across benchmark EO datasets (EuroSAT \cite{helber2019eurosat} and PatternNet \cite{zhou2018patternnet}), both with and without KD from pre-trained ViT models. Remarkably, ResNet-GLUSE achieves over 94\% and 98\% across the evaluation metrics (accuracy, precision, and recall) on the EuroSAT and PatternNet, respectively. Furthermore, it exhibits notable power efficiency, consuming approximately 6 times less energy (13.8 W) than the pre-trained MobileViT (79.23 W) on a GPU. Due to its structural simplicity and computational efficiency, the ResNet-GLUSE model is readily adaptable for neuromorphic Akida edge computing, achieving ultra-low inference power consumption (852.30 mW).

The contributions of this paper are fourfold: 
\begin{itemize} 
\item We propose GLUSE, an adaptive GLU-inspired channel-wise attention mechanism that achieves superior performance over traditional SE blocks. 

\item We comprehensively evaluate the robustness and effectiveness of lightweight ResNet-GLUSE in both standard training and KD scenarios, confirming its advantage. 

\item Leveraging its simplicity and efficiency, ResNet-GLUSE is highly suitable for ultra-low-power neuromorphic onboard satellite deployment. 

\item We further demonstrate that the same GLUSE-enhanced backbone generalizes to image reconstruction under a realistic, end-to-end DVB-S2(X) satellite transmission pipeline with RF impairments and AWGN, confirming its effectiveness on a structurally distinct onboard EO task.
\end{itemize}

\vspace{-2mm}

\section{Related Work}
KD has been widely utilized to boost the performance of lightweight student models, such as ResNet, for onboard satellite processing by transferring knowledge from powerful pre-trained ViT teacher models \cite{le2024semantic}. Although KD significantly enhances the student model’s generalization capabilities, a noticeable performance gap remains between the ResNet student and the large pre-trained ViT teachers, indicating a need for additional architectural improvements. The study \cite{le2024semantic} confirms that even when leveraging semantic information from the teacher model to make student predictions more effective, a considerable gap of approximately 6\% still exists. Alternative strategies, such as increasing the student model’s complexity by making it larger and deeper, have been proposed; however, these approaches come with the drawback of increased computational complexity and, consequently, power consumption.


Alternatively, channel-wise feature recalibration offers a promising strategy for enhancing student model capacity. The SE framework \cite{hu2018squeeze} is a notable example, as it boosts feature expressiveness without adding extra computational overhead. Specifically, SE blocks assign importance weights to each channel, recalibrating channel activations. However, this mechanism is static, applying the same learned attention weights uniformly across all spatial locations. As a consequence, critical features may be suppressed, limiting the model’s adaptability to various spatial contexts \cite{pereira2019adaptive}.

Recent advancements have attempted to enhance SE attention mechanisms. Some studies integrate dense layers into SE blocks to strengthen global context representation \cite{narayanan2024aggregated}. In contrast, others propose lightweight global context modules, such as the Global Context block, to improve feature modeling \cite{cao2019gcnet}. Although these methods improve the learning capacity of lightweight models, they introduce additional computational complexity, making them less suitable for neuromorphic computing, which requires strict architectural simplicity for efficient conversion into spiking neural networks (SNN) \cite{eshraghian2023training, yik2025neurobench}. Furthermore, achieving ultra-low power consumption is essential for satellite-based communication, where energy efficiency is a critical constraint \cite{ortiz2024energy}.

While this study focuses on classification and image reconstruction, object detection and semantic segmentation remain equally important onboard EO tasks, each with its own efficiency-oriented challenges. A recent study provides a comprehensive review of CNN-based object detection and image segmentation methods for EO, surveying and tracing the evolution from early region-proposal detectors to fully convolutional segmentation architectures, while highlighting the persistent trade-off between detection/segmentation accuracy and inference cost in remote sensing applications \cite{hoeser2020object}. At the pixel level, accurate boundary delineation remains a core challenge for segmentation-oriented EO pipelines; recent frequency-domain, boundary-preserving superpixel segmentation approaches address this by leveraging frequency-domain cues to better localize object boundaries prior to or jointly with semantic labeling \cite{zhuang2026frequency}. These complementary lines of work motivate the choice of classification and reconstruction as two representative, resource-constrained tasks in this study, while underscoring detection and segmentation as natural extensions of the proposed GLUSE attention mechanism.

To address these limitations, we propose GLUSE. This GLUSE module retains the computational simplicity of SE while introducing adaptive gating for dynamic channel-wise recalibration. GLUSE balances computational efficiency and enhanced feature learning, making it well-suited for GPU-based execution and edge neuromorphic computing, enabling efficient onboard satellite deployment where power consumption and real-time processing are critical.

The foundation of GLUSE is inspired by GLU \cite{dauphin2017language, le2026transformer}, which selectively emphasizes or suppresses information based on task requirements. Gating mechanisms have been widely adopted in architectures such as Gated Transformer Networks \cite{liu2021gated} and Temporal Fusion Transformers \cite{lim2021temporal}, demonstrating their ability to improve feature selectivity. Mathematically, as described in \cite{lim2021temporal}, a GLU applies a component gating mechanism that modulates the contribution of an input $\eta$:
\vspace{-2mm}
\begin{align}
\text{GLU}_{\omega(\eta)} & = \sigma(W_{1, \omega}\eta + b_{1, \omega}) \odot (W_{2, \omega}\eta + b_{2, \omega} ),
\label{eqn:component_gate}
\end{align}

\noindent where $W_{(.)} $, $b_{(.)}$ are the weights and biases, $\odot$ is the element-wise Hadamard product, and $\sigma(\cdot)$ denotes the sigmoid activation function:
\vspace{-2mm}
\begin{align}
    \sigma(x) = \frac{\exp(x)}{1+\exp(x)}.
\end{align}

GLUs allow selective suppression of irrelevant components by controlling their contribution, effectively skipping nonlinear transformations when necessary, especially enhancing the mutual information between hidden representations, confirmed by \cite{le2026transformer}. By integrating this concept into SE, GLUSE provides a more adaptive and efficient recalibration mechanism.

\section{Proposed Approach}
\begin{figure*}[!ht]
    \centering
    \includegraphics[width=\linewidth]{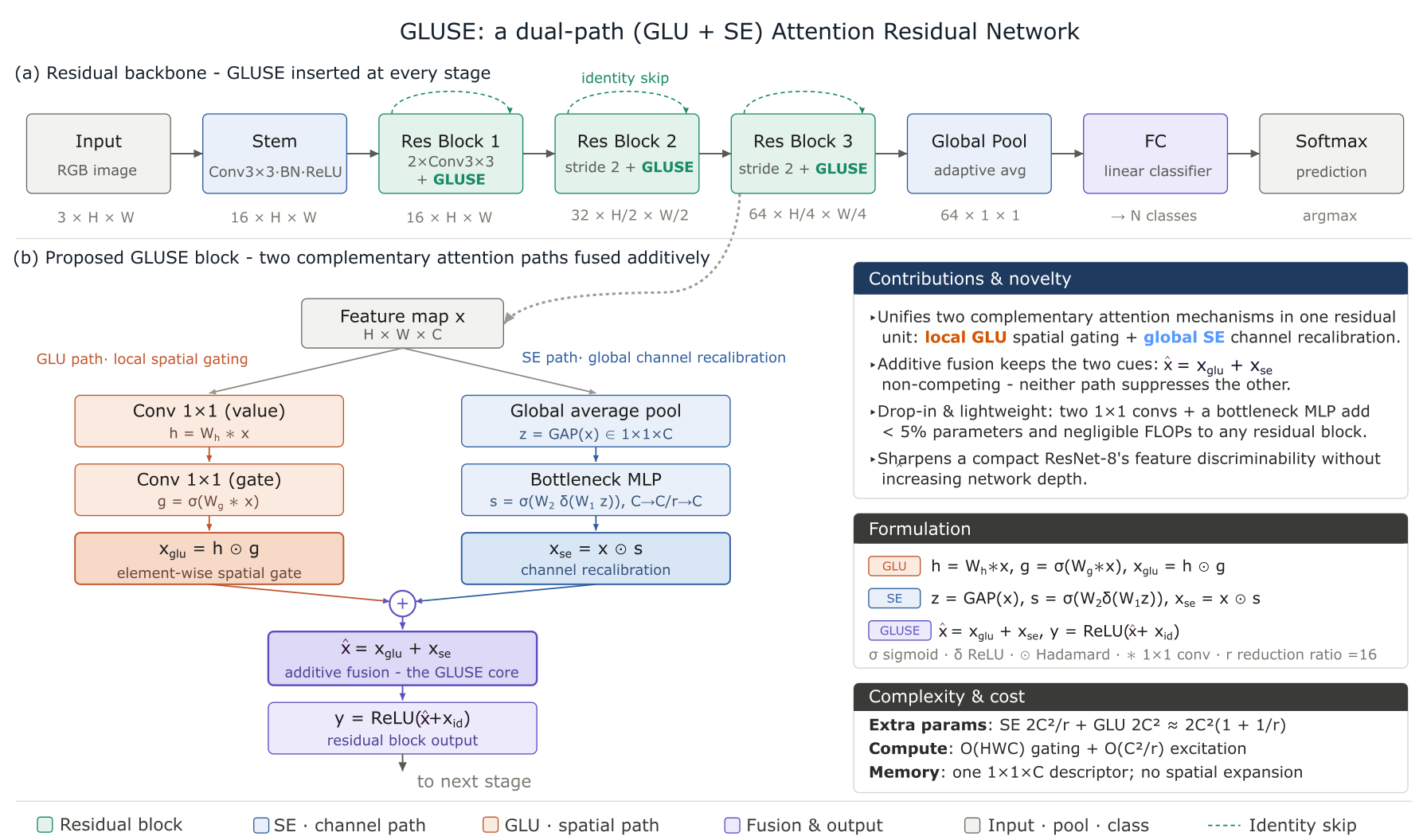}
    \captionsetup{font=small}
    \caption{ResNet-8-GLUSE. (a) The eight-layer residual backbone, with a GLUSE block inserted in every residual stage (channels 16 $\rightarrow$ 32 $\rightarrow$ 64). (b) The proposed GLUSE block runs locally. GLU gating path and a global SE recalibration path in parallel and fuse them additively, $\hat{x} = x_{glu} + x_{se}$, before the residual ReLU - capturing complementary spatial and channel dependencies.}
    \label{fig:GLU_SE_workflow}
    \vspace{-4mm}
\end{figure*}

Let the input feature map be denoted by:
$X \in \mathbb{R}^{H \times W \times C}$,
where $H$, $W$, and $C$ are height, width, and channel dimensions, respectively.

\subsection{Lightweigh ResNet}

As shown in Fig.~\ref{fig:GLU_SE_workflow}(a), the lightweight ResNet variant begins with an initial convolutional layer, followed by three residual blocks that progressively expand the feature channels from 16 to 64. These blocks utilize convolutional layers with varying strides (1 or 2) to balance feature extraction and downsampling, with the proposed GLUSE block embedded directly into each residual unit. The architecture concludes with adaptive average pooling and a fully connected classifier. With its shallow depth, the proposed ResNet in this study is optimized for computationally constrained environments, ensuring efficiency while maintaining representational capacity. Fig.~\ref{fig:GLU_SE_workflow}(b) details the internal structure of the GLUSE block itself. Given an input feature map $\mathbf{x}$, two complementary attention paths operate in parallel: the GLU path performs local spatial gating through two $1{\times}1$ convolutions, $\mathbf{h} = W_h \ast \mathbf{x}$ and $\mathbf{g} = \sigma(W_g \ast \mathbf{x})$, combined via the element-wise gate $\mathbf{x}_\mathrm{glu} = \mathbf{h} \odot \mathbf{g}$; the SE path performs global channel recalibration via global average pooling followed by a bottleneck multi-layer perceptron, yielding the channel-wise scaling factor $\mathbf{s}$ and the recalibrated feature $\mathbf{x}_\mathrm{se} = \mathbf{x} \odot \mathbf{s}$. The two outputs are fused additively as $\hat{\mathbf{x}} = \mathbf{x}_\mathrm{glu} + \mathbf{x}_\mathrm{se}$, ensuring that neither attention mechanism suppresses the other, before being combined with the residual identity path and passed through a ReLU activation. This dual-path design allows GLUSE to capture both fine-grained, spatially localized feature interactions and globally informed channel importance within a lightweight module, adding fewer than 5\% additional parameters and negligible FLOPs relative to the underlying residual block.

\subsection{Squeeze-and-Excitation (SE) Block}
Channel-wise feature recalibration techniques, particularly the SE block, have proven highly effective in enhancing learning models' representational power, including CNN \cite{10433186}, and ResNet \cite{8976424}. SE blocks explicitly model inter-channel dependencies by first applying global average pooling to compress spatial information, followed by two fully connected layers to dynamically recalibrate channel importance. The resultant scalar weights refine feature representations, improving model performance with minimal additional complexity.

The SE block recalibrates channel-wise features through two main steps. First, the \textbf{Squeeze} step to compute the global average pooling (GAP):
\begin{align}
    z_c = \frac{1}{H \times W} \sum_{i=1}^{H}\sum_{j=1}^{W} x_c(i,j),
\quad \mathbf{z}\in\mathbb{R}^{C}.
\end{align}

Then, the \textbf{Excitation} step will perform recalibration through two fully connected convolutional layers:
\begin{align}
    \mathbf{s} = \sigma(W_2(\delta(W_1\mathbf{z}))).
\end{align}

where, $W_1 \in \mathbb{R}^{\frac{C}{r}\times C}$, $W_2 \in \mathbb{R}^{C\times\frac{C}{r}}$, $\delta(\cdot)$ is ReLU activation, $\sigma(\cdot)$ is sigmoid. And, $r$ denotes the reduction ratio used in the SE blocks.

After that, the final recalibration will follow by:
\vspace{-2mm}
\begin{align}
    \tilde{x}_c = s_c \odot x_c,\quad s_c \in (0,1). \label{eq:3}
\end{align}

This recalibration from Eq. \ref{eq:3}  produces channel-wise static recalibration from the SE block.

\vspace{-2mm}

\subsection{Gated Linear Units SE (GLUSE) Block }
Despite SE's effectiveness, their recalibration weights are static during inference, potentially limiting adaptation to varying spatial contexts \cite{pereira2019adaptive}. This study introduces additional adaptive gating operations, enabling recalibration weights to respond dynamically to spatially varying feature distributions.

It is inspired by recent advances in adaptive gating mechanisms, particularly the success of GLU \cite{shazeer2020glu} in capturing complex feature interactions. Unlike traditional gating methods, GLU employs two parallel convolutional pathways: one linear and one gating path, which effectively learn complementary representations, as shown in Fig. \ref{fig:GLU_SE_workflow}. Integrating GLU-inspired gating with the SE framework, the proposed GLUSE approach combines static recalibration from SE with dynamic feature refinement via GLU, enhancing recalibration with greater adaptivity. Initially, GLUSE employs the standard SE procedure. This step provides a channel-wise recalibration that emphasizes informative channels while suppressing redundant ones, precisely as the same SE block from Eq. \ref{eq:3}.
\vspace{-2mm}
\begin{align}
    \mathbf{s} = \sigma(W_2(\delta(W_1\mathbf{z}))), \quad
x_{se} = \mathbf{x}\odot\mathbf{s}.
\end{align}

Then, GLUSE introduces a gating mechanism inspired by GLU to enhance feature refinement adaptivity further. Specifically, two parallel convolutional operations are performed directly on the original feature map $\mathbf{x}$:

Linear transform path:
\vspace{-2mm}
\begin{align}
   \mathbf{h} = W_h*\mathbf{x} 
\end{align}
where $W_h$ is a convolutional kernel.

Gating path:
\vspace{-2mm}
\begin{align}
    \mathbf{g} = \sigma(W_g*\mathbf{x})
\end{align}
where $W_g$ is another convolutional kernel. 

Consequently, this dual-path approach allows GLUSE to dynamically adapt to local feature patterns by generating precise gating masks. Then, the linear path $\mathbf{h}$ and gating path $\mathbf{g}$ outputs are combined via element-wise multiplication, following GLU principles:
\vspace{-2mm}
\begin{align}
    x_{glu} = \mathbf{h}\odot\mathbf{g}.
\end{align}

The GLU operation enables adaptive selection and transformation of features based on their localized importance and relevance, thereby providing richer contextual representation and improved discrimination.

Finally, the SE recalibration output ($x_{se}$) and the adaptive GLU gating output ($x_{glu}$) are summed to form the final enhanced feature representation $\hat{x}$:
\vspace{-2mm}
\begin{align}
   \hat{x} = x_{se} + x_{glu}. 
\end{align}

This final step ensures that both global channel-wise information (from SE) and adaptive local gating (from GLU) collaboratively improve the feature quality.  This approach enhances feature selection by prioritizing significant channels while reducing the influence of less relevant ones. The network can concentrate on critical information by dynamically modulating channel importance, leading to improved feature representation and effective extraction from each channel.

\subsection{Gated SE Block}

To effectively compare and validate the improvements introduced by GLUSE, we also examine the Gated SE block as an intermediate enhancement to the standard SE mechanism. Motivated by the ECA-Net \cite{wang2020eca}, the Gated SE block introduces an additional gating mechanism via a $1 \times 1$ convolution, dynamically recalibrating the weights based on spatial context. This modification enhances SE’s adaptability, making feature modulation more flexible and responsive to input variations.

Improving upon the SE block by adding adaptive gating. First, as in the SE block, the recalibration weights $\mathbf{s}$ are computed as:
\vspace{-2mm}
\begin{align}
    \mathbf{s} = \sigma(W_2(\delta(W_1\mathbf{z}))).
\end{align}

\noindent where $W_1$ and $W_2$ are fully connected layers, $\delta(\cdot)$ is a ReLU activation, and $\sigma(\cdot)$ is the sigmoid function.

 Instead of directly applying $\mathbf{s}$ to scale the feature map $\mathbf{x}$, the Gated SE introduces an adaptive gate $\mathbf{g}$ via a $1 \times 1$ convolution:
 \vspace{-2mm}
\begin{align}
    \mathbf{g} = \sigma(W_g * \mathbf{s}),
\end{align}

\noindent where $W_g$ is a learnable $1 \times 1$ convolution kernel. This additional gating modulates the SE-derived recalibration, allowing dynamic adjustments based on local spatial variations.

Finally, the recalibrated feature map is then obtained by applying the gating function to the original input. This makes the recalibration dynamic, ensuring that channel importance is adjusted contextually rather than being statically assigned. 
\vspace{-2mm}
\begin{align}
    \hat{x} = \mathbf{x} \odot \mathbf{g}.
\end{align}


Table \ref{tab:SE_variants_comp} summarizes the differences between three block SE, Gated SE, and GLUSE. The SE block performs static channel-wise recalibration using GAP followed by fully connected layers. The Gated SE block is an intermediate step between the traditional SE block and the proposed GLUSE mechanism. While Gated SE introduces dynamic gating through a single convolution, it cannot fully adapt feature transformations, as it only modifies the SE weights. The proposed GLUSE further enhances adaptivity by incorporating GLU-inspired gating, utilizing parallel convolutional paths to capture more intricate feature interactions. Consequently, while SE recalibration is static, Gated SE offers moderate adaptability, and GLUSE achieves highly adaptive feature recalibration.

\begin{table*}[!t]
\centering
\footnotesize
\caption{Detailed comparison among SE variants.}
\label{tab:SE_variants_comp}
\begin{tabular}{@{}lccc@{}}
\toprule
\textbf{Operation} & \textbf{SE Block} & \textbf{Gated SE Block} & \textbf{GLUSE} \\ \midrule
Squeeze & $z = \text{GAP}(x)$ & same as SE & same as SE \\[6pt]
Excitation & $\sigma(W_2\delta(W_1 z))$ & same as SE & same as SE \\[6pt]
Gating Mechanism & none & $\sigma(W_g*s)$ & $\mathbf{h}=W_h*x,\;\mathbf{g}=\sigma(W_g*x)$ \\[6pt]
Final Recalibration & $x\odot s$ & $x\odot g$ & $(x\odot s) + (\mathbf{h}\odot \mathbf{g})$ \\[6pt]
Dynamic Gating & No & Yes & Enhanced (GLU-based) \\[6pt]
Feature Interaction & Static & Moderate & Highly adaptive \\ \bottomrule
\end{tabular}
\end{table*}

\begin{table*}[!t]
\centering
\footnotesize
\caption{Computational complexity comparison of plain ResNet with SE, Gated SE, and GLUSE.}
\label{tab:computational_complexity}
\begin{tabular}{@{}l c c@{}}
\toprule
\textbf{Structure} & \textbf{Computational Steps} & \textbf{Complexity (approx.)} \\
\midrule
Plain ResNet & 
Conv: $3\times3\times C \times C$ & 
$\mathcal{O}(HWC^2)$ 
\\[10pt]

SE & 
\begin{tabular}[c]{@{}c@{}}
Global Avg Pool: $\mathcal{O}(HWC)$ \\[2pt]
FC Layers: $\frac{C^2}{r} + \frac{C^2}{r} = \frac{2C^2}{r}$ \\[2pt]
Scaling: $\mathcal{O}(HWC)$
\end{tabular} & 
$\mathcal{O}(HWC + \frac{C^2}{r})$ 
\\[15pt]

Gated SE & 
\begin{tabular}[c]{@{}c@{}}
Gate Conv: $C\times C\times 1\times 1$ \\[2pt]
Gating Operation: $\mathcal{O}(HWC)$
\end{tabular} & 
$\mathcal{O}(HWC + \frac{C^2}{r} + C^2)$
\\[15pt]

GLUSE & 
\begin{tabular}[c]{@{}c@{}}
Linear Conv: $C\times C\times 1\times 1$ \\[2pt]
Gate Conv: $C\times C\times 1\times 1$ \\[2pt]
Gating Operation: $\mathcal{O}(HWC)$ \\[2pt]
Combination (addition): $\mathcal{O}(HWC)$
\end{tabular} & 
$\mathcal{O}(HWC + \frac{C^2}{r} + 2C^2)$ 
\\
\bottomrule
\end{tabular}
\end{table*}

The computational complexity analysis, summarized in Table \ref{tab:computational_complexity}, demonstrates that while the proposed GLUSE slightly increases computational overhead compared to the plain ResNet, SE, and Gated SE blocks, this increase remains modest and manageable. Specifically, the GLUSE structure introduces two additional lightweight convolutional operations (linear and gating paths), resulting in an approximate complexity of $\mathcal{O}(HWC + \frac{C^2}{r} + 2C^2)$, compared to $\mathcal{O}(HWC + \frac{C^2}{r} + C^2)$ for Gated SE, and $\mathcal{O}(HWC + \frac{C^2}{r})$ for standard SE blocks.

\section{Experiments}
\subsection{Datasets}
This study addresses a spectrum of EO tasks, from image classification to image retrieval, while exploring both lower and higher spatial resolutions. To this end, we employ two widely recognized benchmark datasets, EuroSAT \cite{helber2019eurosat} and PatternNet \cite{zhou2018patternnet}, summarized in Table \ref{tab:datasets}. EuroSAT is a Sentinel-2-based benchmark for land use and land cover classification, comprising about 27,000 labeled, georeferenced images at 64x64-pixel resolution across 13 spectral bands. Critically, the images are derived directly from Sentinel-2 multispectral imagery acquired by the European Space Agency's Copernicus program at its native 10 m/pixel ground sampling distance; the 64x64 patch size reflects the spatial extent of the sampled scene rather than a downsampled or synthetically reduced version of a larger image. EuroSAT is divided into 10 classes and provides a balanced dataset for onboard deep learning methods for real-time tasks such as environmental monitoring and precision agriculture. In contrast, PatternNet targets image retrieval using 30,400 high-resolution images (256x256 pixels) spanning 38 categories, constructed from real remote sensing imagery with spatial resolutions ranging from 0.116 to 4.693 m/pixel, substantially finer than those of typical onboard EO sensors targeted in this study. This diverse coverage, spanning both coarse- and fine-resolution genuine satellite and aerial imagery, supports the development of retrieval-focused approaches to tackle more complex, fine-grained RS challenges, while also providing a realistic basis for evaluating onboard models at sensor-representative spatial resolutions.
\vspace{-2mm}

\begin{table*}[!t]
\centering
\caption{Comparison of two satellite imagery datasets for EO oriented task.}
\footnotesize
\begin{tabular}{lcccccccc}
\hline
\textbf{Dataset} & \textbf{Classes} & \textbf{Images/Class} & \textbf{Images} & \textbf{Train (70\%)} & \textbf{Test (30\%)} & \textbf{Resolution (m)} & \textbf{Size} & \textbf{Task} \\
\hline
EuroSAT \cite{helber2019eurosat}   & 10 & 2000 - 3000 & 27,000 & 18,900 & 8,100 & 0.3           & 64$\times$64   & Classification \\
PatternNet \cite{zhou2018patternnet} & 38 & 800        & 30,400 & 21,280 & 9,120 & 0.116 - 4.693  & 256$\times$256 & Retrieval       \\
\hline
\end{tabular}
\label{tab:datasets}
\end{table*}

\subsection{Training Strategy}

To validate the effectiveness and robustness of GLUSE, we evaluate its performance under two distinct training scenarios:  
(1) \textbf{Standard training}, where ResNet-GLUSE is trained conventionally without additional supervision, and  
(2) \textbf{Dual-teacher KD training}, where knowledge is transferred from large pretrained ViTs to enhance the model's generalization. 

Standard training, in this approach, ResNet-GLUSE is trained directly using conventional supervised learning. While this setup ensures a fair baseline comparison against existing SE and Gated SE models, its performance is inherently limited by the available training data and model capacity. With KD training, to further enhance model performance while maintaining computational efficiency for onboard satellite deployment, we employ KD - a technique where a compact student model learns from larger, more expressive teacher models \cite{papa2024survey}. Proposed initially to reduce deep learning models’ computational burden \cite{hinton2015distilling}, KD has since evolved as a powerful method for enabling lightweight models to acquire complex representations while achieving competitive accuracy \cite{stanton2021does}.

\begin{figure*}[!ht]
    \centering
    \includegraphics[scale=0.425]{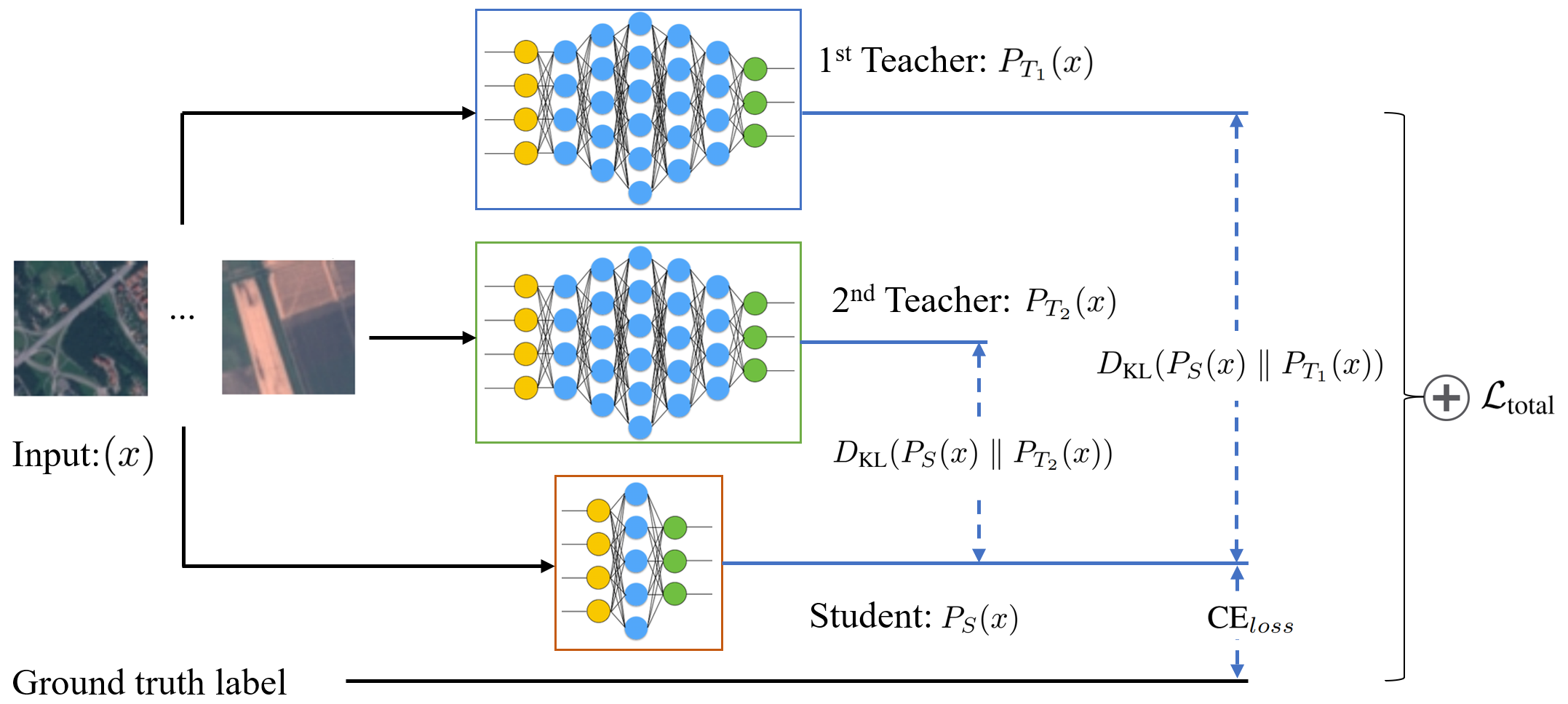}
    \captionsetup{font=small}
    \caption{The schematic workflow of dynamic weighting in dual-teacher KD.}
    \label{fig:workflow}
    \vspace{-4mm}
\end{figure*} 

In this study, we adopt dual-teacher KD to distill semantic knowledge from ViTs into ResNet-GLUSE. Traditional KD methods, which enforce strict alignment with a single teacher’s predictions via Kullback-Leibler (KL) divergence \cite{van2014renyi}, often suffer from training instability and suboptimal performance when the teacher model is uncertain \cite{huang2022knowledge}. To mitigate these limitations, we introduce dynamic weighting in dual-teacher KD (Algorithm \ref{alg:proposed_dual_teacher_distillation}), where the weight assigned to each teacher is adjusted based on confidence scores. This adaptive strategy enables the student to prioritize the more reliable knowledge source, improving generalization across data distributions.

As shown in Fig. \ref{fig:workflow}, given an input $x$, the semantic distillation process starts by computing softened probability distributions for the teacher models and the student model. This is achieved by scaling their logits with a temperature parameter $\tau$. For teacher model $T_1$, the softened probability distribution is:
\begin{equation}
    P_{T_1}(x) = \text{softmax}\left(\frac{T_1(x)}{\tau}\right),
\end{equation}
and similarly for teacher model $T_2$:
\begin{equation}
    P_{T_2}(x) = \text{softmax}\left(\frac{T_2(x)}{\tau}\right),
\end{equation}
with the student model $S$:
\begin{equation}
    P_S(x) = \text{softmax}\left(\frac{S(x)}{\tau}\right).
\end{equation}

\noindent Confidence for each teacher is computed as the average of the maximum probabilities in their softened distributions:
\begin{equation}
    C_{T_1} = \mathbb{E}\left[\max(P_{T_1}(x))\right], \quad C_{T_2} = \mathbb{E}\left[\max(P_{T_2}(x))\right].
\end{equation}
Based on these confidence scores, we dynamically adjust the weights $\alpha$ and $\beta$ assigned to each teacher in the distillation loss $\text{KD}_{\text{loss}}$. If both confidence scores are significantly below a predefined threshold $\delta$, both teachers are ignored ($\alpha = \beta = 0$). If either confidence score is close to the threshold, we prioritize the more reliable teacher by reducing the weight of the less reliable one, with minimum weights set by $w_{min}$. When both teachers are above the threshold, equal weights ($\alpha = \beta = 0.5$) are used.

The distillation loss $\text{KD}_{loss}$, a weighted KL divergence between the student’s and each teacher’s softened probabilities, is then computed, with the weights $\alpha$ and $\beta$ reflecting each teacher’s confidence.
\vspace{-5mm}

\begin{equation}
    \text{KD}_{\text{loss}} = \alpha \cdot D_{\text{KL}}(P_S(x) \parallel P_{T_1}(x)) + \beta \cdot D_{\text{KL}}(P_S(x) \parallel P_{T_2}(x)),
\end{equation}

Where the KL divergence $D_{\text{KL}}$ for each teacher-student pair is scaled by the temperature squared, $\tau^2$, to stabilize training:

\begin{equation}
    D_{\text{KL}}(P_S(x) \parallel P_{T_i}(x)) = \frac{1}{\tau^2} \sum_{j} P_{T_i}(x)_j \log\left(\frac{P_{T_i}(x)_j}{P_S(x)_j}\right)
\end{equation}

The total distillation loss is calculated as a combination of the classification loss, $\text{CE}_{\text{loss}}$, and the distillation loss $\text{KD}_{\text{loss}}$. A classification loss $\text{CE}_{loss}$ between the student’s predictions and the true labels is calculated to ground the student’s learning in teacher guidance and actual labels, where:

\begin{equation}
    \text{CE}_{\text{loss}} = - \sum_{i} y_i \log \left( P_S(x)_i \right),
\end{equation}

Then, the final combined loss, $\mathcal{L}_{\text{total}}$, integrates these components: a weighted combination of the $\text{CE}_{loss}$ and the $\text{KD}_{loss}$. This framework allows the student to leverage insights from both teachers selectively, focusing on the most reliable sources for improved generalization and adaptability across instances during training.

\begin{equation}
    \mathcal{L}_{\text{total}} = \left(1 - \frac{\alpha + \beta}{2}\right) \cdot \text{CE}_{\text{loss}} + \frac{\alpha + \beta}{2} \cdot \text{KD}_{\text{loss}}
\end{equation}

Recent work \cite{le2024board, nguyen2025semantic} has identified EfficientViT and MobileViT as the two most effective ViTs for EO image classification tasks. Accordingly, we select EfficientViT \cite{liu2023efficientvit} and MobileViT \cite{MehtaR23} as our teacher models. By combining ResNet-GLUSE with dynamic dual-teacher KD, we aim to maximize accuracy and resource efficiency.

\begin{algorithm}[htbp]
\caption{Dynamic Weighting Dual-Teacher KD}
\label{alg:proposed_dual_teacher_distillation}
\footnotesize
\begin{algorithmic}[1]
\REQUIRE Input batch $(x,y)$, student model $S$, teacher models $T_1$ and $T_2$, temperature $\tau$, confidence threshold $\delta$, minimum weight $w_{min}$
\ENSURE Combined loss $\mathcal{L}_{total}$ for backpropagation

\STATE \textbf{Forward Pass:}
\STATE \quad Compute logits: $l_S \leftarrow S(x)$, $l_{T1} \leftarrow T_1(x)$, $l_{T2} \leftarrow T_2(x)$
\STATE \quad Compute softened predictions using temperature $\tau$:
\STATE \quad\quad $P_S \leftarrow \text{softmax}(l_S/\tau)$, $P_{T1} \leftarrow \text{softmax}(l_{T1}/\tau)$, $P_{T2} \leftarrow \text{softmax}(l_{T2}/\tau)$

\STATE \textbf{Compute Teacher Confidence Scores:}
\STATE \quad $C_{T1} \leftarrow \text{mean}(\max(P_{T1}))$, $C_{T2} \leftarrow \text{mean}(\max(P_{T2}))$

\STATE \textbf{Dynamic Weight Adjustment:}
\IF{$C_{T1}, C_{T2} < 0.4$}
    \STATE $\alpha, \beta \leftarrow 0.0, 0.0$ \hfill // Both teachers ignored
\ELSIF{$C_{T1} < \delta$ \textbf{and} $C_{T2} < \delta$}
    \STATE $\alpha \leftarrow \max(0.5-(\delta - C_{T1}), w_{min})$
    \STATE $\beta \leftarrow \max(0.5-(\delta - C_{T2}), w_{min})$
\ELSIF{$C_{T1} < \delta$}
    \STATE $\alpha, \beta \leftarrow 0.3, 0.7$ \hfill // Prioritize confident Teacher 2
\ELSIF{$C_{T2} < \delta$}
    \STATE $\alpha, \beta \leftarrow 0.7, 0.3$ \hfill // Prioritize confident Teacher 1
\ELSE
    \STATE $\alpha, \beta \leftarrow 0.5, 0.5$ \hfill // Equal weighting
\ENDIF

\STATE \textbf{Compute Knowledge Distillation (KD) Loss:}
\STATE \quad $\text{KD}_{loss} \leftarrow \tau^2 \left[\alpha \cdot D_{KL}(P_S \parallel P_{T1}) + \beta \cdot D_{KL}(P_S \parallel P_{T2})\right]$

\STATE \textbf{Compute Classification Loss (Cross-Entropy):}
\STATE \quad $\text{CE}_{loss} \leftarrow -\sum_{i} y_i \log(P_S)_i$

\STATE \textbf{Combine Losses:}
\STATE \quad $w \leftarrow \frac{\alpha+\beta}{2}$
\STATE \quad $\mathcal{L}_{total} \leftarrow (1-w)\text{CE}_{loss} + w\text{KD}_{loss}$

\STATE \textbf{Backpropagate} $\mathcal{L}_{total}$

\end{algorithmic}
\normalsize
\end{algorithm}

\vspace{-2mm}

\subsection{Implementation}

\begin{table}[!t]
\centering
\footnotesize
\caption{Experiment parameters setting}
\vspace{-2mm}
\begin{tabular}{|l|c|}
\hline
\textbf{Parameter} & \textbf{Value} \\
\hline
Batch size & 64 \\
Optimizer & AdamW \\
Learning rate & 0.00025 \\
Weight decay & 0.0005 \\
Scheduler & ReduceLRonPlateau \\
Threshold ($\delta$) & 0.6 \\
Temperature ($\tau$) & 5 \\
Min weight ($w_{min}$) & 0.1 \\
\hline
\end{tabular}
\label{table:experiment_parameters}
\vspace{-4mm}
\end{table}

All experiments are conducted on the NVIDIA RTX™ 6000 GPU, 48 GB GDDR6. Experiments were implemented using the Scikit-learn library \cite{scikit-learn} and PyTorch. For inference, we also tested the proposed model on the in-lab Akida BrainChip edge neuromorphic computing platform. The data was split into 70\% for training and 30\% for testing. In addition, we applied Batch Normalization \cite{bjorck2018understanding} to improve model stability. The data transformation pipeline applies standard preprocessing steps to ensure compatibility with various neural network backbones \cite{cubuk2019autoaugment}. First, it resizes images to 64×64 pixels, converts them to tensors, and normalizes pixel values using the mean \([0.485, 0.456, 0.406]\) and standard deviation \([0.229, 0.224, 0.225]\), which aligns with the input requirements of widely used pretrained models, including ViTs and ResNets. Finally, the detailed parameters are summarized in Table \ref{table:experiment_parameters}.

To comprehensively evaluate the performance of our multiclass classification model across classes, we employ three key metrics: accuracy, precision, and recall (sensitivity) \cite{goutte2005probabilistic}. These metrics are calculated for each class individually and then aggregated using macro-averaging to assess the model's performance as follows:
\vspace{-2mm}
\begin{align}
&\text{Accuracy} = \sum_{k=1}^K\frac{\text{TP}_k}{N}  \\ 
&\text{Precision} = \frac{1}{N} \sum_{i=1}^{K} N_k \frac{\text{TP}_k}{\text{TP}_k + \text{FP}_k}  \\ 
&\text{Recall} = \frac{1}{N} \sum_{i=1}^{K} N_k \frac{\text{TP}_k}{\text{TP}_k + \text{FN}_k} 
\end{align}
where $N$ is the total number of data points across all classes. $K$ is the total number of classes. $N_k$ is the number of data points in class $k$. $TP_k$ is True Positives, $FP_k$ is False Positives, $FN_k$ is False Negatives for class $k$, respectively. We use weighted precision and recall to ensure that each class is given equal importance, thereby providing a balanced evaluation of the model’s classification capabilities across the entire dataset. These macro-averaged evaluation metrics will select the best models in the final analysis.

\section{Results and Discussion}
\label{sec:results_discussion}

\subsection{Image Classification}
\label{subsec:classification}

\begin{figure*}[!ht]
    \centering
    \includegraphics[scale=0.45]{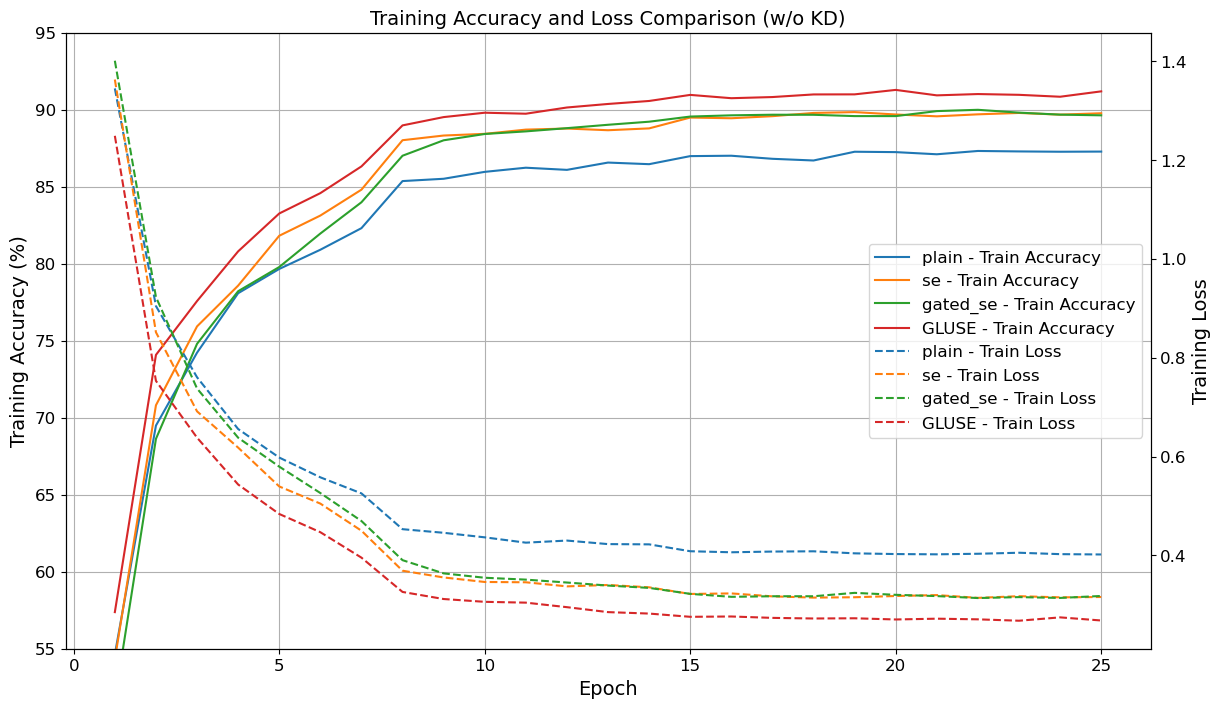}
    \includegraphics[scale=0.45]{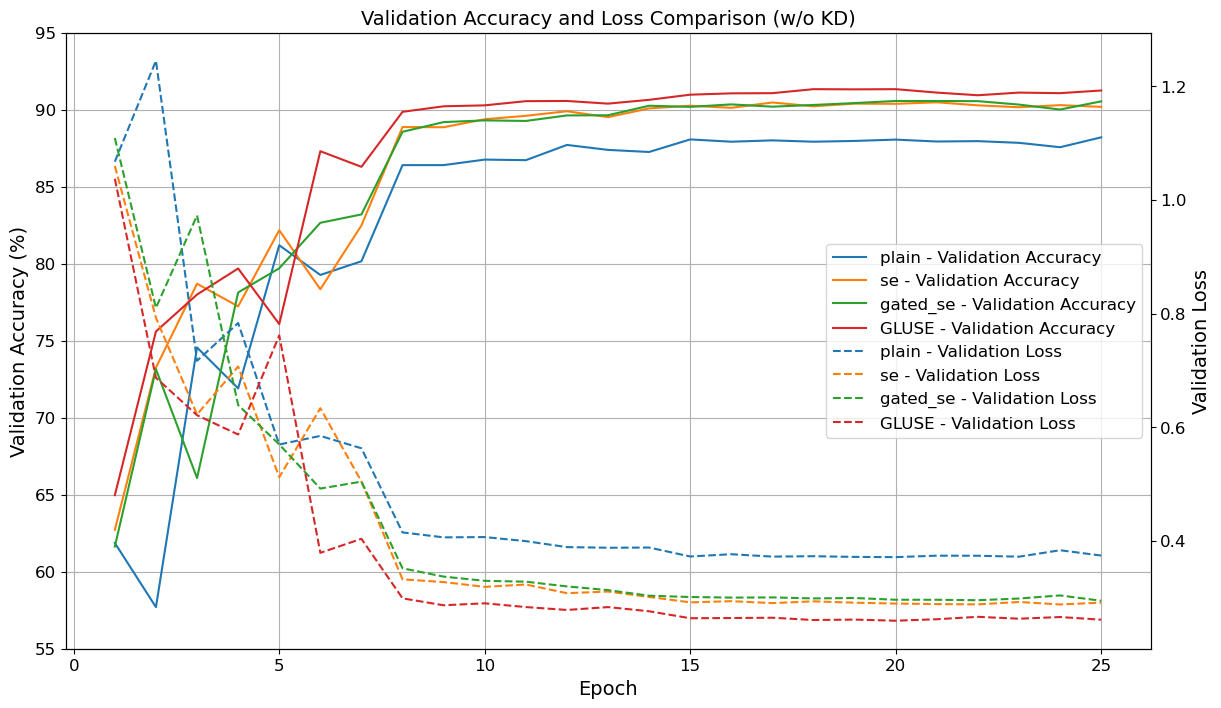}
    \captionsetup{font=small}
    \caption{Training (Top) and validation (Bottom) learning curve from standard training strategy.}
    \label{fig:woKD_performance}
    \vspace{-2mm}
\end{figure*} 

\begin{figure*}[!t]
    \centering
    \includegraphics[scale=0.45]{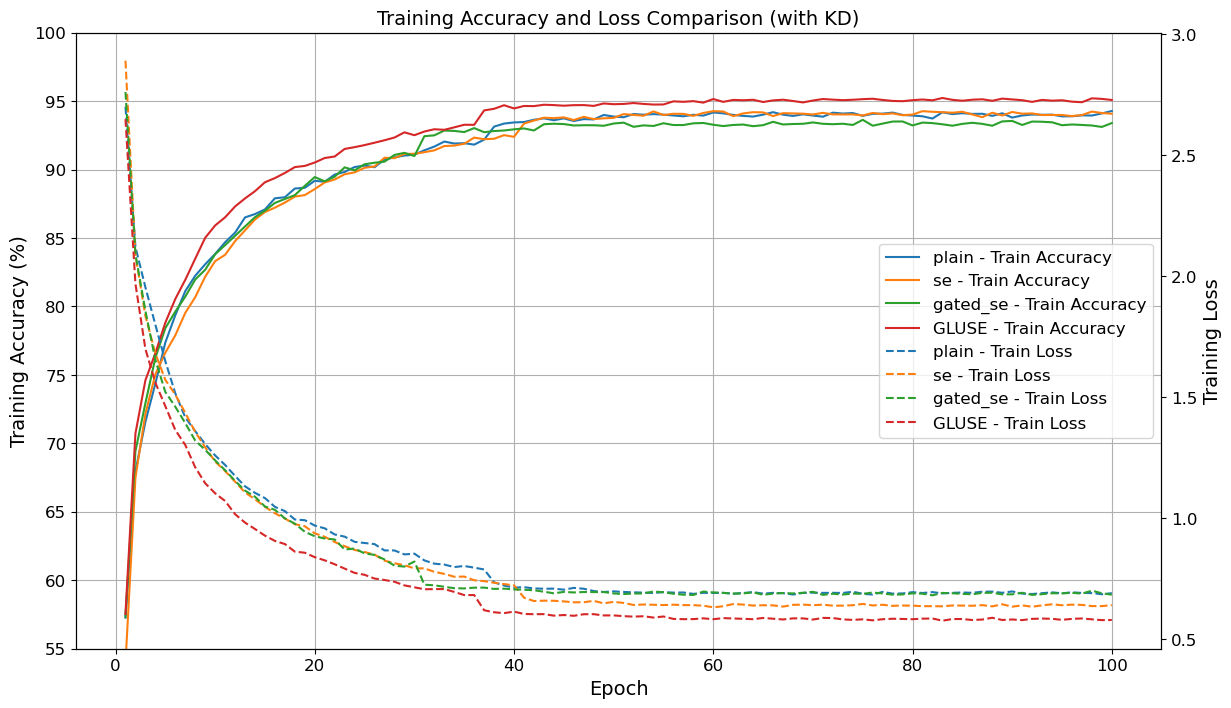}
    \includegraphics[scale=0.45]{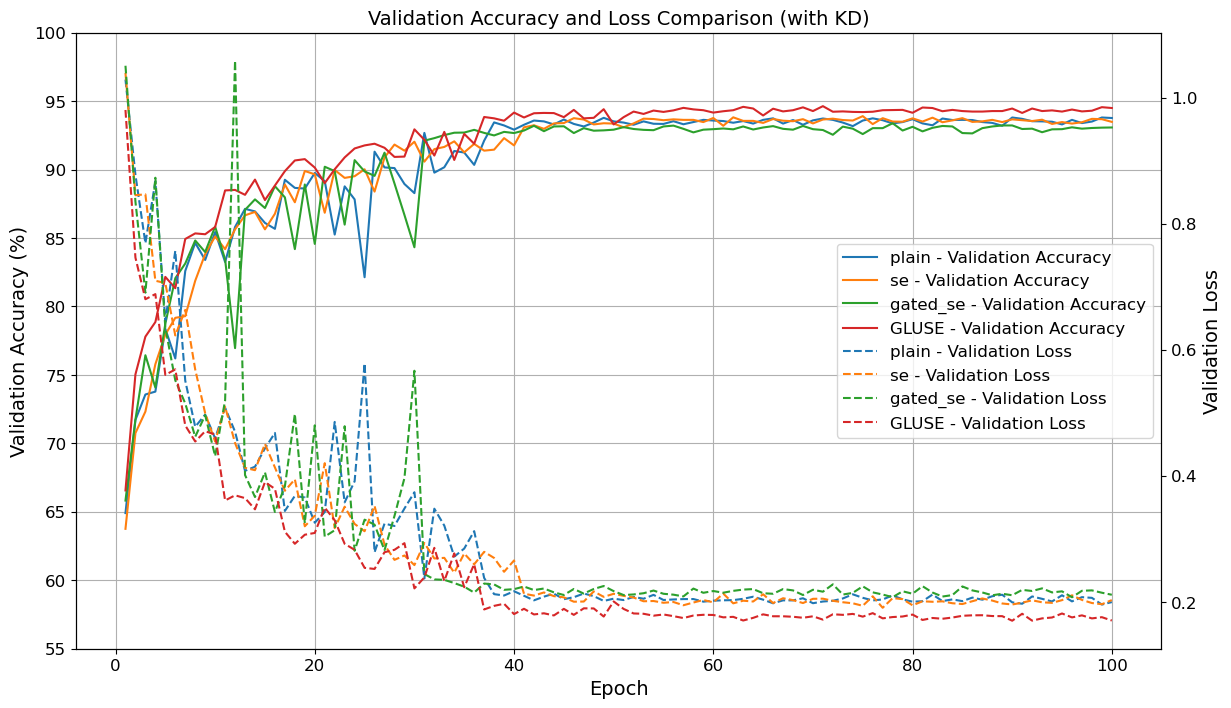}
    \captionsetup{font=small}
    \caption{Training (Top) and validation (Bottom) learning curve with KD training strategy.}
    \label{fig:wKD_performance}
    \vspace{-2mm}
\end{figure*} 

The experimental learning curves from Fig. \ref{fig:woKD_performance}, and \ref{fig:wKD_performance} demonstrate that the proposed ResNet-GLUSE consistently outperforms the plain ResNet, ResNet-SE, and ResNet-gated-SE architectures, both with and without KD.  Without KD, ResNet-GLUSE achieves faster convergence, higher validation accuracy ($\approx$ 91\%), and lower validation loss, highlighting the efficacy of the adaptive gating mechanism in feature recalibration. With KD, all models significantly improve; however, ResNet-GLUSE shows notably superior performance, quickly converging and maintaining validation accuracy around 94-95\%, surpassing other models by a clear margin. The consistently lowest loss values further confirm that GLUSE effectively leverages the distilled knowledge from large pretrained ViT teachers. Additionally, it is vital to note that an early stopping strategy was employed to mitigate overfitting, revealing that training without KD converged after 25 epochs, while with KD, achieving stable convergence required 100 epochs.


\begin{table}[ht]
\centering
\footnotesize
\caption{Comparison of model performance on EuroSat.}
\vspace{-2mm}
\begin{tabular}{lccc}
\hline
\textbf{Models} & \textbf{Accuracy} $(\uparrow)$ & \textbf{Precision} $(\uparrow)$ & \textbf{Recall} $(\uparrow)$\\
\hline
\multicolumn{4}{c}{\textbf{Pretrained Models}} \\
EfficientViT & 98.76 & 98.77 & 98.76 \\
MobileViT     & \textbf{99.09} & \textbf{99.09} & \textbf{99.09} \\
\hline
\multicolumn{4}{c}{\textbf{w/o KD}} \\
ResNet                   & 87.76 & 87.70 & 87.76 \\
ResNet-SE                & 90.54 & 90.50 & 90.54 \\
ResNet-gated-SE          & 90.19 & 90.19 & 90.19 \\
ResNet-GLUSE & 91.05 & 91.1 & 91.05 \\
\hline
\multicolumn{4}{c}{\textbf{with KD}} \\
ResNet                   & 92.88 & 93.07 & 92.88 \\
ResNet-SE                & 93.49 & 93.47 & 93.49 \\
ResNet-gated-SE          & 93.07 & 93.04 & 93.07 \\
ResNet-GLUSE & \underline{\textbf{94.63}} & \underline{\textbf{94.61}} & \underline{\textbf{94.63}} \\
\hline
\end{tabular}
\label{tab:EuroSat_performance}
\vspace{1mm} 
\captionsetup{font=footnotesize}
\caption*{\hspace{-4.9cm}\textbf{Bold} denotes the best values.\\
         \hspace{-2.4cm}\underline{\textbf{Bold and underline}} denote the second best values.}
\vspace{-6mm}
\end{table}

\begin{table}[ht]
\centering
\footnotesize
\caption{Comparison of model performance on PatternNet.}
\vspace{-2mm}
\begin{tabular}{lccc}
\hline
\textbf{Models} & \textbf{Accuracy} $(\uparrow)$ & \textbf{Precision} $(\uparrow)$ & \textbf{Recall} $(\uparrow)$ \\
\hline
EfficientViT & 99.52 & 99.52 & 99.52 \\
MobileViT     & \textbf{99.66} & \textbf{99.66} & \textbf{99.66} \\
\hline
\multicolumn{4}{c}{\textbf{w/o KD}} \\
ResNet                   & 80.18 & 79.42 & 80.18 \\
ResNet-SE                & 86.34 & 86 & 86.34 \\
ResNet-gated-SE          & 86.02 & 85.80 & 86.02 \\
ResNet-GLUSE & 88.16 & 87.93 & 88.16 \\
\hline
\multicolumn{4}{c}{\textbf{with KD}} \\
ResNet                   & 97.73 & 97.70 & 97.73 \\
ResNet-SE                & 98.02 & 98 & 98.02 \\
ResNet-gated-SE          & 97.98 & 97.98 & 97.98 \\
ResNet-GLUSE & \underline{\textbf{98.09}} & \underline{\textbf{98.09}} & \underline{\textbf{98.09}} \\
\hline
\end{tabular}
\label{tab:PatternNet_performance}
\vspace{1mm} 
\captionsetup{font=footnotesize}
\caption*{\hspace{-4.9cm}\textbf{Bold} denotes the best values.\\
         \hspace{-2.4cm}\underline{\textbf{Bold and underline}} denote the second best values.}
\end{table}

Furthermore, across the EuroSat and PatternNet datasets, as shown in Table \ref{tab:EuroSat_performance} and \ref{tab:PatternNet_performance}, the proposed ResNet-GLUSE consistently outperforms the standard ResNet, ResNet-SE, and ResNet Gated SE architectures. Specifically, without KD, the ResNet-GLUSE achieves higher accuracy, precision, and recall, demonstrating that the adaptive gating mechanism inspired by GLU significantly enhances the traditional SE framework by providing more effective channel-wise feature recalibration. When leveraging KD, ResNet-GLUSE further boosts its performance substantially, achieving metrics closely comparable to the top-performing, with only a slight dip in accuracy. Particularly on PatternNet, ResNet-GLUSE reaches 98.09\% accuracy (99.66\% for MobileViT), and on EuroSat, it attains 94.63\% accuracy (99.09\% for MobileViT).

\begin{table*}[h!]
\centering
\footnotesize
\caption{Model Comparison on Parameters, FLOPs, Size, Inference Time, and Power Consumption}
\vspace{-2mm}
\label{tab:model_comparison}
\begin{tabular}{|l|c|c|c|c|c|}
\hline
\textbf{Models} & \textbf{Total Parameters} ($\downarrow$) & \textbf{FLOPs} ($\downarrow$) & \textbf{Size (MB)} ($\downarrow$) & \textbf{Inference time (s)} ($\downarrow$) & \textbf{Power (W)} ($\downarrow$) \\ \hline
ResNet8 & \textbf{98,522} & \textbf{60,113,536} & \textbf{5.95} & \textbf{5.84} & \textbf{10.94 $\pm$ 0.83} \\ \hline
ResNet8-SE & 120,589 & 60,151,840 & 6.04 & 5.87 & 11.51 $\pm$ 1.58 \\ \hline
ResNet8-gated-SE & 126,077 & 60,271,904 & 6.06 & 5.99 & 13.05 $\pm$ 0.74 \\ \hline
ResNet8-GLUSE & 131,565 & 66,557,984 & 7.92 & 6.01 & 13.80 $\pm$ 1.39 \\ \hline
EfficientViT  & 3,964,804 & 203,533,056 & 38.19 & 10 & 29.04 $\pm$ 0.96 \\ \hline
MobileViT  & 4,393,971 & 1,843,303,424 & 259.30 & 16 & 79.23 $\pm$ 1.45 \\ \hline
\end{tabular}
\label{tab:complexity_compa}
\vspace{1.5mm} 
\captionsetup{font=footnotesize}
\caption*{\hspace{-11.5cm}\textbf{Bold} denotes the best values.}
\vspace{-6mm}
\end{table*}

As shown in Table \ref{tab:model_comparison}, the proposed ResNet8-GLUSE exhibits remarkable computational efficiency compared to large pretrained ViT-based models (EfficientViT and MobileViT). Although ResNet8-GLUSE is slightly larger and more computationally demanding than the plain ResNet8, ResNet8-SE, and ResNet8-gated-SE variants (with parameters at 131,565 and FLOPs at 66.56M), it remains significantly simpler, smaller, and more efficient than MobileViT (4.39M parameters, 1.84G FLOPs). Specifically, ResNet8-GLUSE achieves approximately 33 times fewer parameters and 27 times fewer FLOPs than MobileViT, translating into substantially lower inference power consumption (13.80W vs. 79.23W). These results underline that ResNet-GLUSE, especially combined with KD, can achieve competitive performance levels while offering significant computational efficiency, making it highly suitable for onboard satellite image classification and retrieval.

It is worth noting that the power summarization in Table \ref{tab:model_comparison}, measured as instantaneous GPU power draw via NVML (nvidia-smi), does not scale proportionally with parameter count or FLOPs. At this model scale (98K-132K parameters), inference is dominated by kernel-launch and scheduling overhead rather than sustained computation, making power draw more sensitive to kernel concurrency than to total arithmetic work. GLUSE's dual-path design, which executes the SE and GLU branches as two independent parallel sub-paths rather than the single sequential pathway used by SE and Gated-SE, issues a larger number of concurrent kernel launches per forward pass, which elevates instantaneous power disproportionately relative to its modest increase in parameters (+33.5\%) and inference time (+2.9\%). This is consistent with the larger power variance observed for ResNet8-GLUSE ($\pm$ 1.39 W) relative to plain ResNet8 ($\pm$ 0.83 W). Despite this increase, GLUSE's absolute power consumption remains nearly 6x times lower than MobileViT (13.80 W vs. 79.23 W), preserving its efficiency advantage for onboard deployment.

\begin{figure}[!ht]
    \centering
    \includegraphics[scale=0.75]{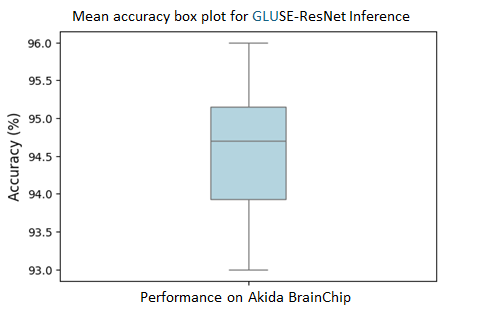}
    \captionsetup{font=small}
    \caption{Performance of the ResNet-GLUSE during the inference on Akida BrainChip neuromorphic computing hardware.}
    \label{fig:akida_inference}
    \vspace{-2mm}
\end{figure} 

More importantly, due to its architectural simplicity and compactness, the ResNet-GLUSE model can easily be mimicked, adapted, and deployed on the Akida Brainchip neuromorphic computing platform \cite{akida}. Experimental deployment on Akida hardware further highlights its exceptional efficiency, with an extremely low inference power consumption averaging 877 mW, achieving an inference energy consumption of just 182.42 mJ/frame, and maintaining a practical frame rate of 4.81 fps. Such results highlight the potential of the ResNet-GLUSE model to operate efficiently on neuromorphic hardware, enabling energy-efficient onboard image analysis in resource-constrained environments. From the boxplot accuracy distribution, depicted in Fig. \ref{fig:akida_inference}, the model’s accuracy ranges from about 93\% to nearly 96\%, with a median of approximately 94.7\%, demonstrating stable and high performance across multiple runs, further confirming its robust inference performance under varied operational conditions.

\begin{figure*}[!h]
    \centering
    \includegraphics[scale=0.395]{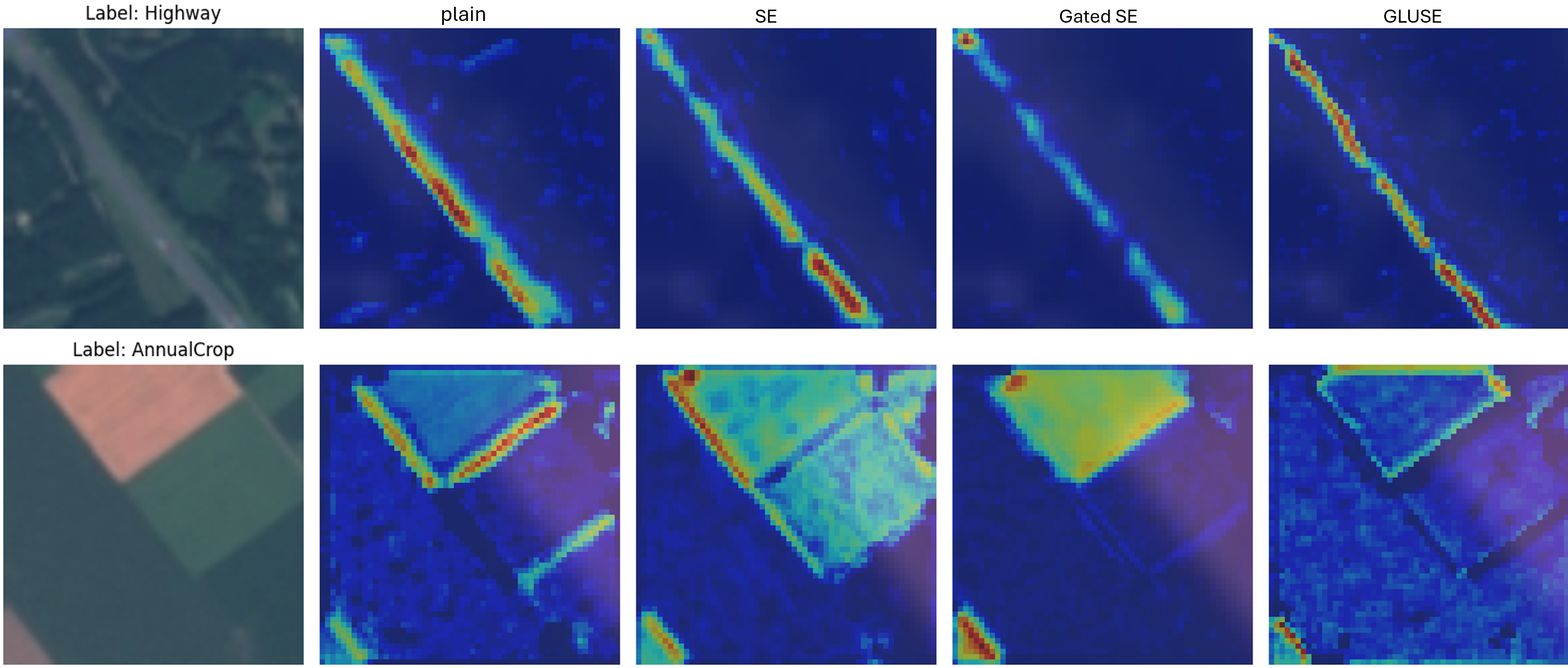}
    \includegraphics[scale=0.395]{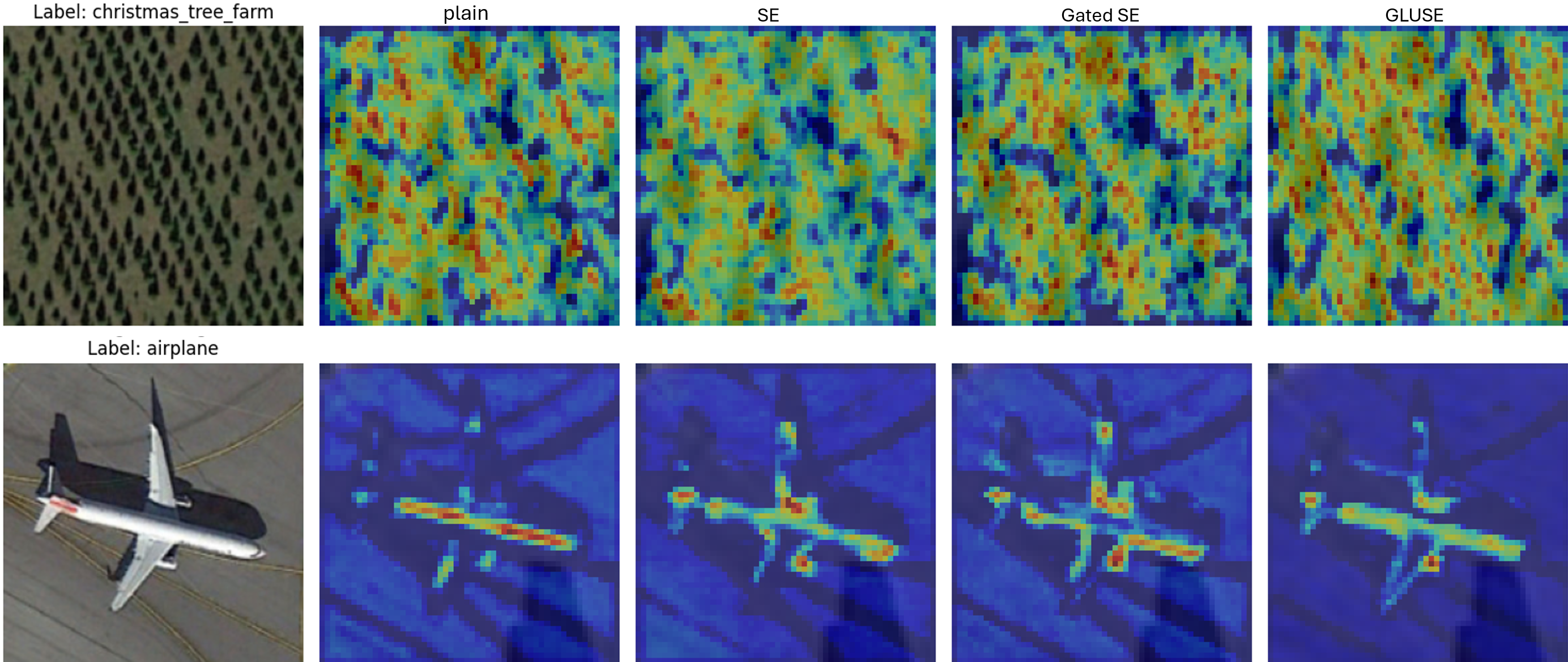}
    \captionsetup{font=small}
    \caption{Qualitative results of different channel-wise attention for the EuroSAT, and PatternNet data by using Grad-CAM.}
    \label{fig:EuroSat_CAM}
    \vspace{-4mm}
\end{figure*}

Furthermore, the Grad-CAM \cite{selvaraju2017grad} visualization qualitatively confirms the effectiveness of the proposed ResNet-GLUSE compared to plain ResNet, SE, and Gated-SE models, as shown in Fig. \ref{fig:EuroSat_CAM}. Across diverse classes (``highway," ``annual crop," ``airplane," and particularly ``christmas tree farm"), GLUSE consistently produces more transparent and more precise activation maps aligned closely with the ground truth. Notably, in the ``christmas tree farm" scenario, the GLUSE model distinctly captures individual tree distributions, highlighting fine-grained features clearly, while other methods yield significantly noisier and less structured activations. The adaptive gating mechanism within GLUSE effectively recalibrates channel-wise attention, enhancing the interpretability and specificity of the learned features. This qualitative evidence further validates GLUSE’s ability to identify relevant features accurately. This hybrid approach from GLU and SE significantly enhances adaptivity and recalibration precision, improving representation clarity and classification.

\begin{figure*}[!t]
    \centering
    \includegraphics[scale=0.525]{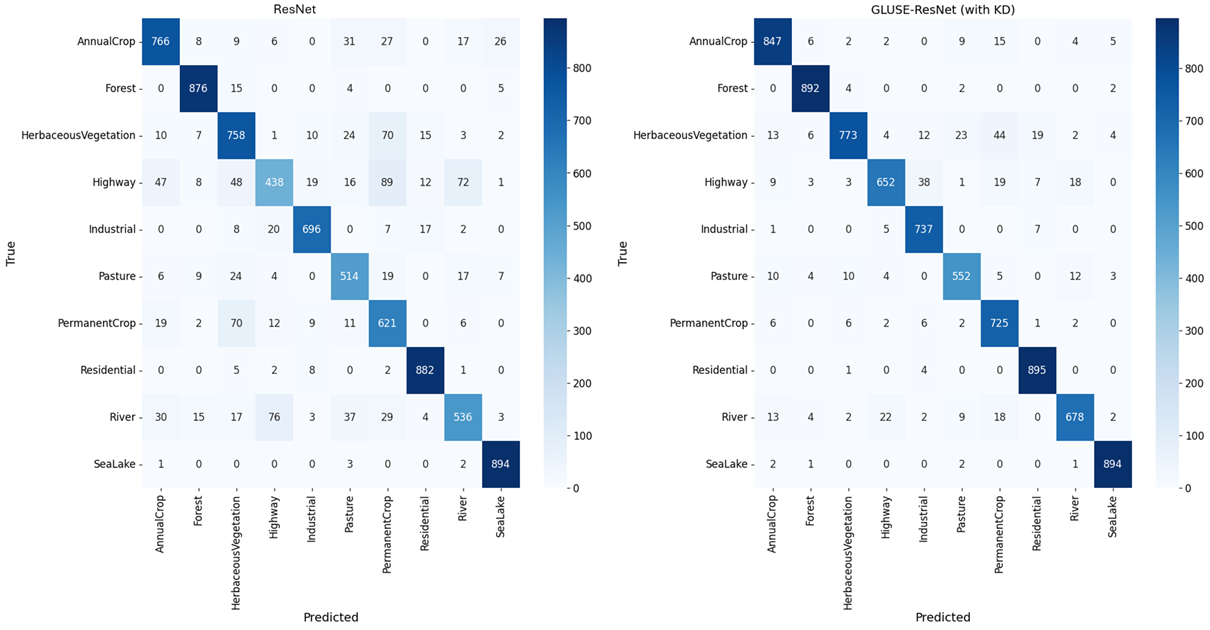}
    \captionsetup{font=small}
    \caption{Confusion matrix from ResNet (left) and ResNet-GLUSE with KD (right) on the EuroSat dataset.}
    \label{fig:conf_matrix_eurosat}
    \vspace{-3mm}
\end{figure*}

\begin{figure*}[!t]
    \centering
    \includegraphics[width=\linewidth]{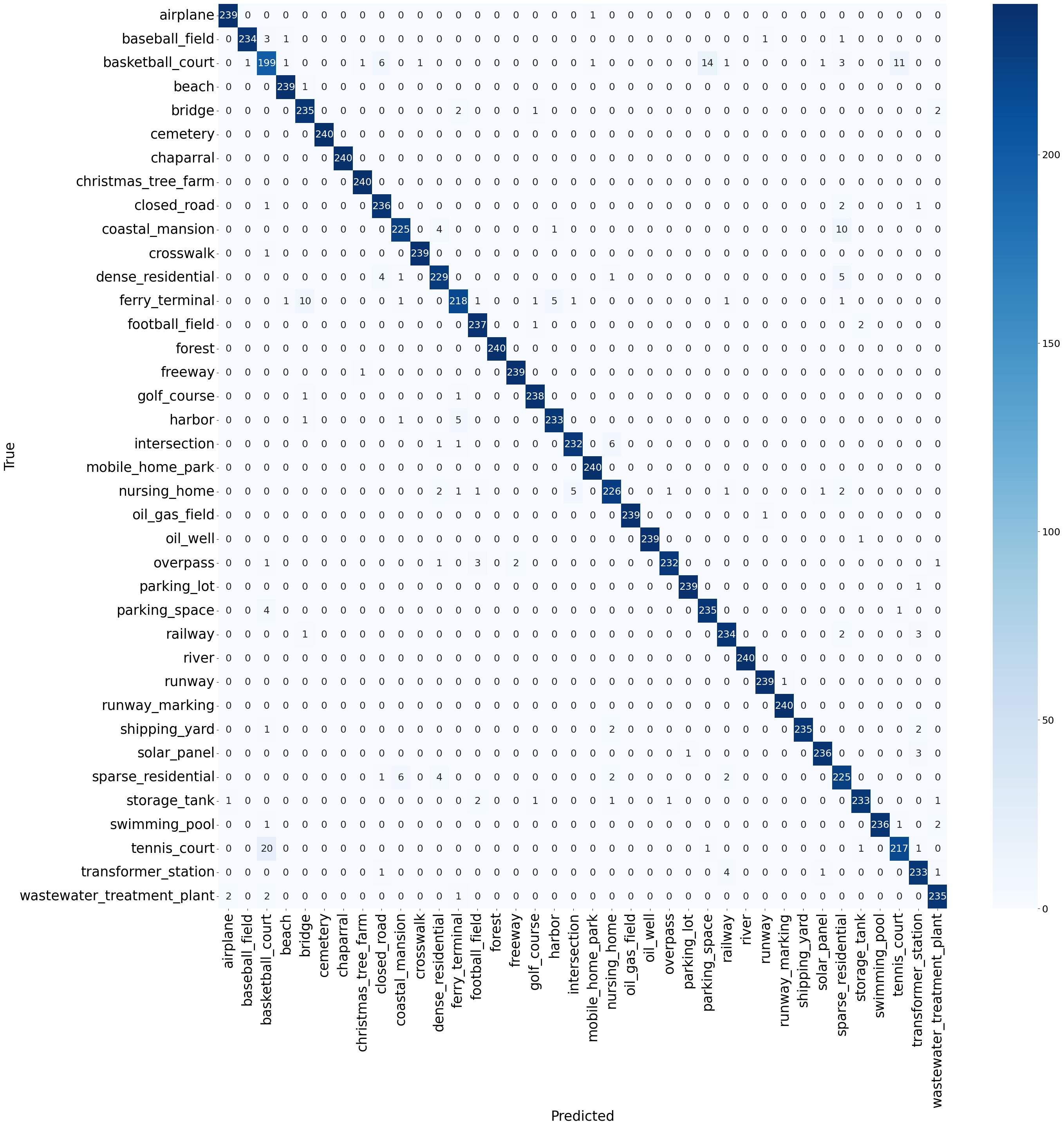}
    \captionsetup{font=small}
    \caption{Confusion matrix from ResNet-GLUSE with KD on the PatternNet dataset.}
    \label{fig:conf_matrix_Pattern}
\end{figure*}

Lastly, comparing the diagonal entries in the two confusion matrices reveals that ResNet-GLUSE produces $\approx$ 8\% higher overall accuracy than the baseline ResNet, translating to an increase of roughly 660+ correct predictions on this 9,000-image test set, shown in Fig. \ref{fig:conf_matrix_eurosat}. Notably, performance in challenging classes such as Highway improves substantially, with correct classifications rising from 438 to 652, and River also sees a significant jump from 536 to 678. In contrast, classes already well-modeled by the baseline (e.g., SeaLake) show similar high performance under ResNet-GLUSE. These results underscore the model’s capacity to discriminate more effectively among visually similar categories, boosting both recall and overall accuracy. A similar trend is also observed in the PatternNet dataset, where ResNet-GLUSE achieves outstanding performance across nearly all classes, illustrated in Fig. \ref{fig:conf_matrix_Pattern}. From the PatternNet confusion matrix, each class is predicted with near-perfect accuracy—for instance, airplane achieves 239 correct predictions (99.6\%) out of 240 samples. By combining knowledge distillation with a more capable student architecture, ResNet-GLUSE demonstrates clear and consistent gains across a wide range of EO classes.

\begin{table*}[!t]
\centering
\caption{Comparison with AiTLAS results \cite{dimitrovski2023current} on EuroSAT and PatternNet.
}
\label{tab:literature_comparison}
\footnotesize
\begin{tabular}{llcc}
\hline
\textbf{Dataset} & \textbf{Model} & \textbf{Params} & \textbf{Accuracy (\%)} \\
\hline
\multirow{4}{*}{EuroSAT}
  & AiTLAS, best from-scratch (EfficientNetB0) & 5.2M  & 97.80 \\
  & AiTLAS, best pretrained (SwinT)            & 49.7M & 98.94 \\
  & ResNet8-GLUSE, w/o KD (ours)               & 132K  & 91.05 \\
  & ResNet8-GLUSE, with KD (ours)              & 132K  & 94.63 \\
\hline
\multirow{4}{*}{PatternNet}
  & AiTLAS, best from-scratch (DenseNet161)    & 26.4M & 99.24 \\
  & AiTLAS, best pretrained (DenseNet161)      & 26.4M & 99.74 \\
  & ResNet8-GLUSE, w/o KD (ours)               & 132K  & 88.16 \\
  & ResNet8-GLUSE, with KD (ours)              & 132K  & 98.09 \\
\hline
\end{tabular}
\end{table*}

Table~\ref{tab:literature_comparison} compares ResNet8-GLUSE against results reported by AiTLAS \cite{dimitrovski2023current} for the same two datasets. While their best pretrained models achieve marginally higher accuracy on EuroSAT (98.94\%) and PatternNet (99.74\%), they rely on architectures with 200 - 380$\times$ more parameters than ResNet8-GLUSE (132K). This confirms that near-ceiling accuracy on these benchmarks is attainable only at substantially higher model complexity, reinforcing the accuracy-efficiency trade-off targeted by this study for onboard satellite deployment.

\subsection{Image Reconstruction}
\label{subsec:reconstruction}
 
Beyond classification accuracy, a deployed onboard EO pipeline must also contend with the physical transmission of the (possibly compressed) image data from the satellite to the ground station, where channel impairments can substantially distort the recovered bitstream before any downstream task - classification, retrieval, or visual inspection - is performed. Our earlier work~\cite{nguyen2025semantic, chou2025semantic, nguyen2026resource, nguyen2026semantic} introduced a semantic-loss modeling framework that quantifies how source coding (compression) and transmission losses jointly degrade task-oriented accuracy in EO applications, and evaluated EfficientViT, MobileViT, ResNet50-DINO, and ResNet8-KD under varying channel conditions and compression ratios. That study established that the two loss sources interact non-trivially: neither the compression level nor the channel signal-to-noise ratio (SNR) alone determines end-task performance.
 
Building on this insight, we extend the present study to directly evaluate whether the proposed ResNet-GLUSE architecture, originally designed for onboard classification, is similarly effective at \emph{image reconstruction} when used as a denoising autoencoder at the ground segment. This complements the classification results in Section~\ref{sec:results_discussion} by examining a second, equally critical onboard EO task category, and provides a more realistic, standards-compliant channel model than the static benchmark datasets alone can offer, directly addressing the practical relevance of the proposed architecture under physical-layer satellite transmission impairments.
 
\subsubsection{Transmission pipeline}

We simulate the full satellite-to-ground link using the second-generation Digital Video Broadcasting, satellite (DVB-S2X) standard, the de facto standard for broadband satellite communication. As illustrated in \cite{nguyen2025semantic}, each EuroSAT and PatternNet image is first compressed to a target quality level, then encoded into the DVB-S2(X) bitstream at the transmitter. The signal is then subjected to realistic RF impairments before passing through an additive white Gaussian noise (AWGN) channel parameterized by the energy-per-symbol-to-noise ratio ($E_s/N_0$). At the ground station, the DVB-S2(X) receiver demodulates and decodes the bitstream, yielding a \emph{recovered compressed image} that reflects the cumulative degradation from both lossy compression and channel noise. ResNet-GLUSE is then applied as a reconstruction network, trained to map the recovered compressed image back to its original, uncompressed counterpart.
 
\subsubsection{Experimental design}
For both EuroSAT and PatternNet, we sweep two independent factors: (i) the compression quality, from 10\% to 100\% in steps of 10\%, and (ii) the channel condition, $E_s/N_0 \in \{1, 2, 3, 4\}$~dB, covering the transition from a noise-dominated to a compression-dominated regime. Reconstruction fidelity is evaluated using three standard image-quality metrics \cite{gao2009image} computed between the reconstructed and original images: peak signal-to-noise ratio (PSNR), structural similarity index (SSIM), and mean squared error (MSE).

\begin{figure*}[!ht]
    \centering
    \includegraphics[width=\linewidth]{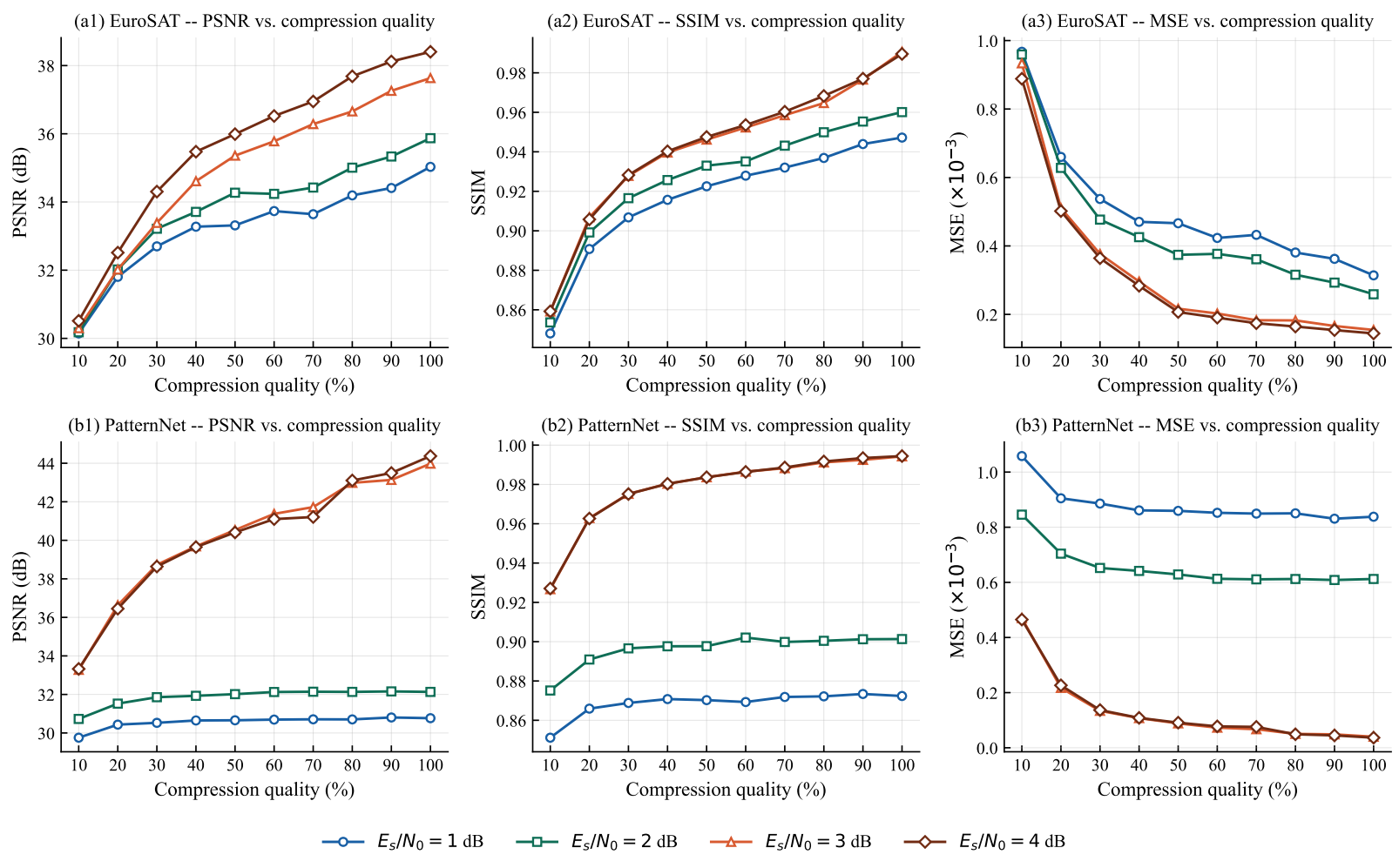}
    \captionsetup{font=small}
    \caption{Reconstruction quality of compressed EO images after transmission through the simulated DVB-S2(X) satellite link (RF impairments + AWGN) at $E_s/N_0 \in \{1,2,3,4\}$ dB across compression-quality levels (10\%--100\%) on (a1 - a3) EuroSAT and (b1 - b3) PatternNet. PSNR, SSIM, and MSE show an SNR threshold around 3~dB, above which reconstruction fidelity scales with compression quality.}
  \label{fig:reconstruction_quality}
    \vspace{-4mm}
\end{figure*} 

Fig.~\ref{fig:reconstruction_quality} quantifies the reconstruction fidelity of ResNet-GLUSE across the full sweep of compression quality and channel SNR for both EuroSAT and PatternNet. At low channel quality ($E_s/N_0 \leq 2$~dB), all three metrics are tightly clustered and largely insensitive to compression level: PSNR remains confined to a narrow band (e.g., 30 - 38~dB for EuroSAT, 30 - 44~dB for PatternNet), and SSIM and MSE exhibit comparably flat trends, indicating that channel noise, rather than compression, is the dominant source of reconstruction error in this regime. Once $E_s/N_0$ reaches 3--4~dB, however, a clear separation emerges: PSNR rises by 6--12~dB relative to the low-SNR curves, SSIM approaches 0.99, and MSE drops by an order of magnitude, with all three metrics now scaling consistently with compression quality rather than remaining flat. This transition forms a distinct waterfall pattern that is consistent across both datasets, despite their differing spatial resolution and scene complexity, indicating that the behavior reflects a property of the channel-compression interaction rather than a dataset-specific artifact. These results confirm that ResNet-GLUSE reliably reconstructs EO imagery once a modest SNR threshold is met, while also revealing that, below this threshold, gains in compression quality provide limited benefit unless paired with a corresponding improvement in channel conditions.

\begin{figure*}[!ht]
    \centering
    \includegraphics[scale=0.39]{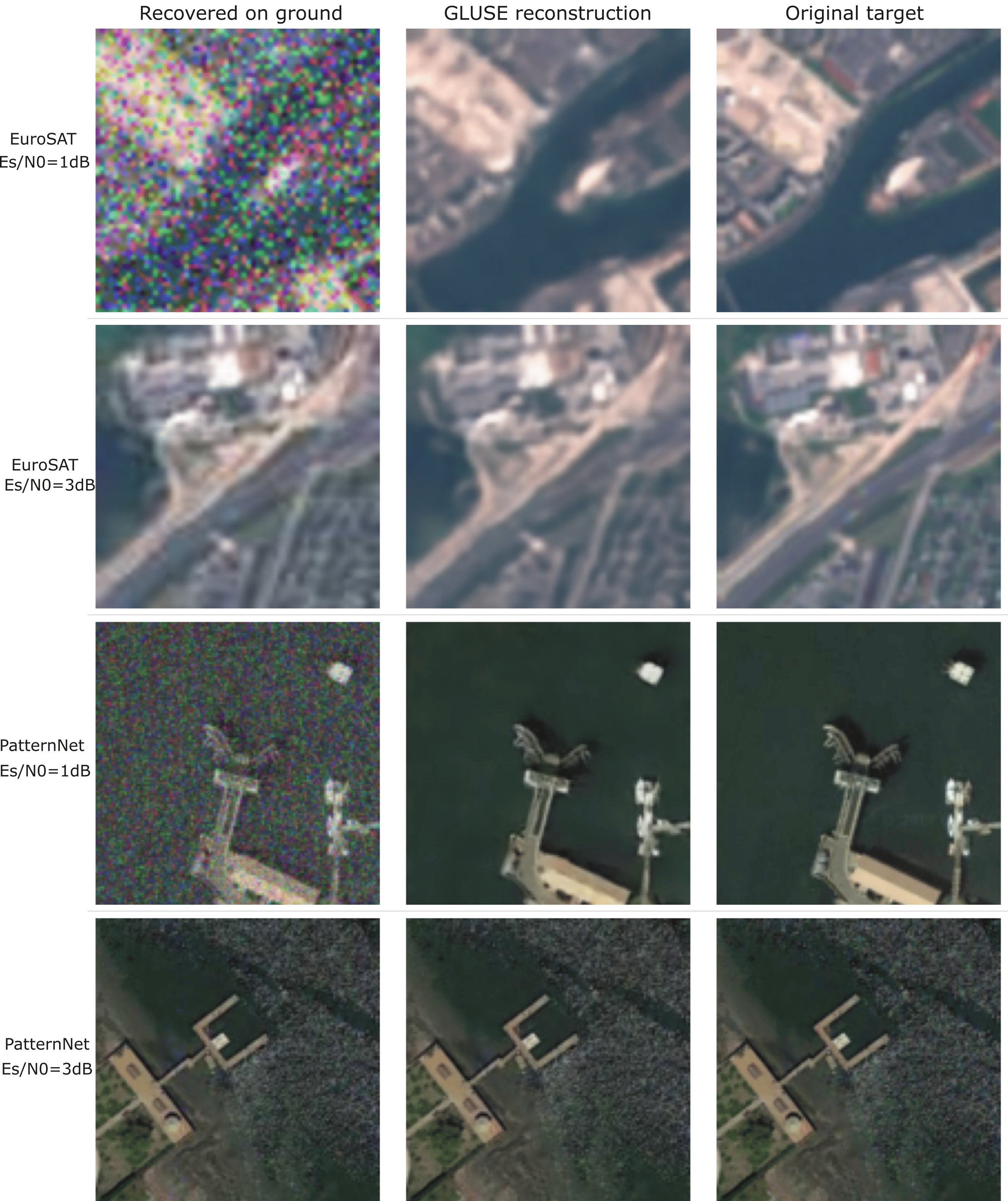}
    \captionsetup{font=small}
    \caption{Qualitative reconstruction examples under the simulated DVB-S2(X) satellite link (RF impairments + AWGN) at 50\% compression quality.   Each row shows the image recovered on the ground (left, corrupted by channel noise), the corresponding GLUSE reconstruction (middle),  and the original uncompressed image (right), for EuroSAT and PatternNet at $E_s/N_0 = 1$~dB (noise-dominated regime) and $3$~dB (compression-dominated regime). At 1~dB, the recovered image is visually unrecognizable, yet ResNet-GLUSE restores the main scene structure; at 3~dB, the reconstruction is nearly indistinguishable from the original.}
  \label{fig:qualitative_comparison}
    \vspace{-4mm}
\end{figure*} 

Fig.~\ref{fig:qualitative_comparison} provides qualitative confirmation of the SNR-dependent reconstruction behavior observed in the quantitative metrics. At $E_s/N_0 = 1$~dB, the image recovered on the ground is dominated by dense channel noise, with the original scene content, a river bend in EuroSAT and a ferry-terminal pier structure in PatternNet, rendered visually unrecognizable; nevertheless, ResNet-GLUSE successfully reconstructs the principal scene layout, correctly recovering the river course, field boundaries, and the pier's crane-like structure, despite having no access to a clean reference at inference time. At $E_s/N_0 = 3$~dB, the recovered image is already largely legible, and the reconstruction becomes nearly indistinguishable from the original target, preserving fine-grained details such as building edges, water texture, and small floating objects that are easily lost under heavier compression or noise. This qualitative contrast between the two channel conditions visually corroborates the waterfall effect identified in Fig.~\ref{fig:reconstruction_quality}: below the SNR threshold, ResNet-GLUSE operates as a denoising network restoring scene structure from severely corrupted input, whereas above it, the network performs fine-detail refinement closer to a conventional compression-artifact removal task. Overall, these examples demonstrate that the proposed architecture remains a competent reconstruction backbone across both extremes of channel quality, supporting its suitability for the variable and often unfavorable link conditions encountered in onboard satellite-to-ground transmission.

\section{Limitations and Future Work}
\label{sec:limitations}
For future work, we will extend ResNet-GLUSE to semantic segmentation and object detection tasks, enabling its applicability to more complex EO vision-based problems. Additionally, we plan to integrate multimodal data, including synthetic aperture radar (SAR) and hyperspectral imaging, to further enhance its robustness and generalization in diverse satellite-based remote sensing applications.

A further limitation of this study is its reliance on EuroSAT and PatternNet, which, while derived from genuine Sentinel-2 and high-resolution aerial/satellite sensors, respectively, are nonetheless bound to 64x64 and 256x256 pixel patches. Evaluating ResNet-GLUSE on higher-resolution satellite imagery, such as full-scene Sentinel-2 tiles or very-high-resolution commercial sensor products, would provide stronger evidence of its scalability to larger spatial extents and more complex scene content. We identify this as an important direction for future work, building on recent super-resolution image segmentation and reconstruction \cite{zhuang2026frequency, hassan2025exploring}, which offer a promising route toward extending the proposed attention mechanism to higher-resolution imagery, a task beyond classification.

Another limitation relates to the power overhead introduced by GLUSE's dual-path design. Because the SE and GLU branches execute as two independent parallel sub-paths rather than the single sequential pathway used by SE and Gated-SE, GLUSE issues more concurrent kernel launches per forward pass, which disproportionately elevates instantaneous GPU power draw relative to its modest increase in parameters and inference time. While this overhead remains small in absolute terms and does not offset GLUSE's overall efficiency advantage over larger ViT-based models, it suggests that further architectural or kernel-fusion optimizations, e.g., merging the SE and GLU sub-paths into a single fused kernel, could reduce this power gap and improve energy efficiency on resource-constrained onboard hardware. We leave such implementation-level optimization to future work.

\section{Conclusions}

This study presents GLUSE, an adaptive channel-wise calibration mechanism that enhances lightweight ResNet models for onboard EO image classification. Extensive experiments on EuroSAT and PatternNet confirm that ResNet-GLUSE outperforms SE and Gated SE models, consistently achieving over 94\% accuracy, precision, and recall across diverse training scenarios. With KD, it approaches ViT-level performance, achieving over 98\% accuracy while maintaining a significantly lower computational footprint.

It is worth emphasizing that GLUSE's contribution is most pronounced in the standard training setting without KD, where it improves EuroSAT accuracy from 87.76\% to 91.05\% (+3.29 points over plain ResNet), compared to a smaller +1.75-point gain with KD; this is expected, as KD from pretrained ViT teachers already injects substantial external semantic information, compressing the headroom available to any architectural change to the student backbone. On PatternNet, where the KD-assisted baseline already exceeds 97.7\% accuracy, the remaining margin for improvement is similarly constrained regardless of the attention mechanism used. Importantly, this modest accuracy gain comes at negligible additional cost: as shown in Table \ref{tab:complexity_compa}, GLUSE adds only a small fraction of the parameters and FLOPs of ResNet8 and remains two orders of magnitude smaller than MobileViT, so that even a one-point gain represents a favorable accuracy-complexity trade-off for onboard deployment. Beyond classification accuracy, the Grad-CAM visualizations in Fig. \ref{fig:EuroSat_CAM} show that GLUSE produces qualitatively cleaner and more spatially precise activation maps than SE and Gated-SE, most notably resolving individual tree structures in the "Christmas tree farm" class, where the baselines yield noisy, unstructured activations, indicating a more interpretable feature representation beyond what accuracy alone conveys. Finally, Section \ref{subsec:reconstruction} demonstrates that the same GLUSE-enhanced backbone is also an effective image-reconstruction network under realistic DVB-S2(X) satellite transmission conditions, a second, structurally distinct EO task in which the benefit of the proposed attention mechanism is not constrained by the near-ceiling KD performance observed in classification.

ResNet-GLUSE optimizes accuracy, efficiency, and resource consumption. It reduces energy consumption by up to 6× compared to MobileViT on GPUs and enables ultra-low-power inference (852.30 mW) on the Akida Brainchip. Despite a slight increase in complexity over SE, its adaptive gating mechanism delivers substantial performance gains with minimal overhead. Future work will extend ResNet-GLUSE to higher-resolution satellite imagery and to object detection and semantic segmentation tasks, explore kernel-fusion optimizations to reduce the power overhead introduced by its dual-path design, and further validate the reconstruction pipeline under additional channel conditions and compression standards, as detailed in Section \ref{sec:limitations}.

\section*{Acknowledgment}
We acknowledge Dr. Geoffrey Eappen for his valuable efforts in mimicking and implementing the model on Akida. This work was funded by the Luxembourg National Research Fund (FNR), with the granted SENTRY project corresponding to grant reference C23/IS/18073708/SENTRY. 

\bibliographystyle{IEEEtran}
\bibliography{IEEEabrv,Bibliography}

\makeatletter
\def\@IEEEBIOskipN{0.25\baselineskip}
\def\@IEEEBIOphotowidth{0.8in}      
\def\@IEEEBIOphotodepth{0.95in}     
\def\@IEEEBIOhangwidth{0.94in}      
\def\@IEEEBIOhangdepth{0.95in}      
\makeatother
\linespread{0.9}

\begin{IEEEbiography}[{\includegraphics[width=0.8in, height=0.95in, clip, keepaspectratio]{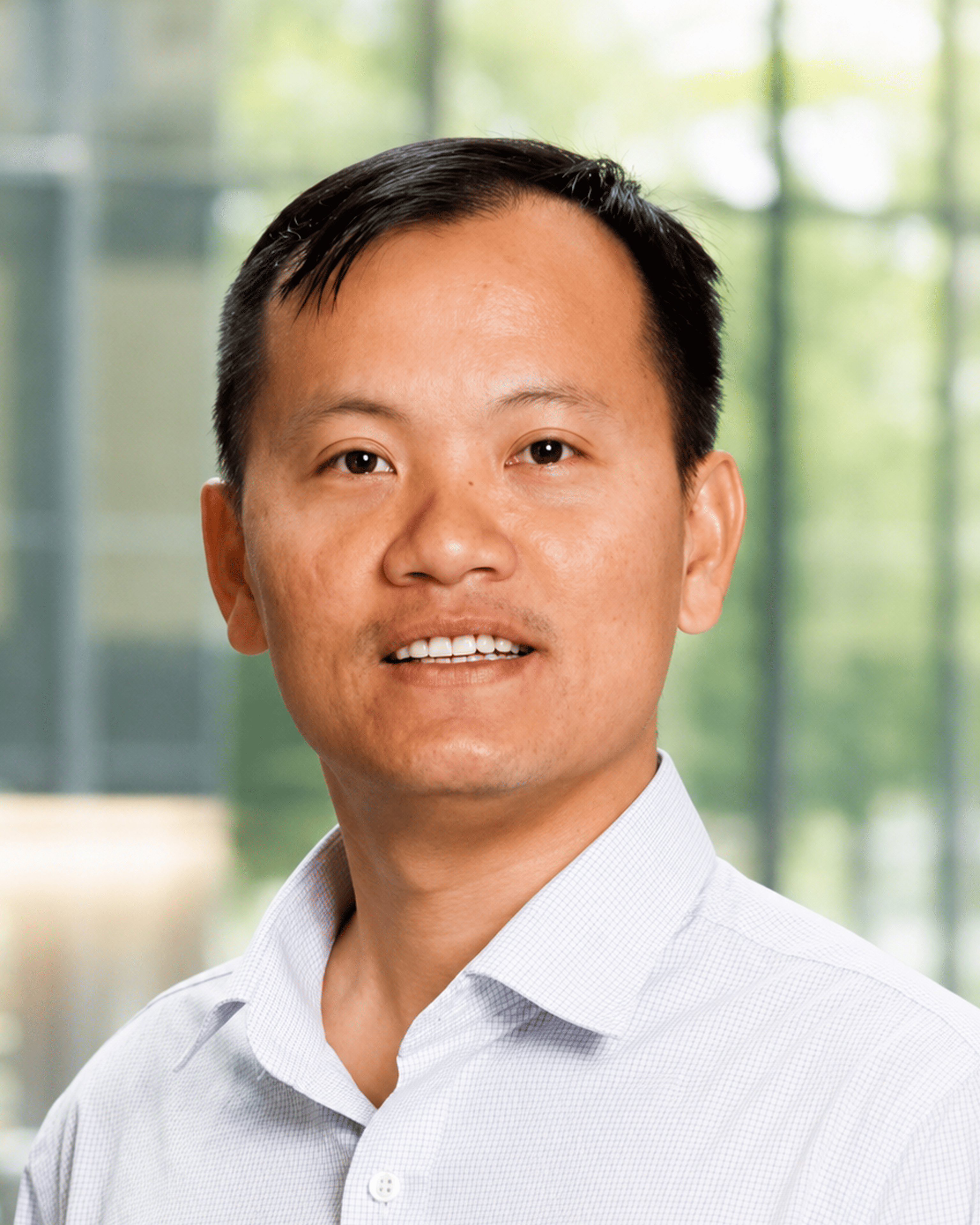}}]{Thanh-Dung Le} (Senior Member, IEEE) received a B.Eng. degree from Can Tho University, Vietnam, an M.Eng. degree from Jeju National University, S. Korea, and a Ph.D. degree in engineering from École de Technologie Supérieure, University of Québec, Canada. From 2023 to 2024, he was a Postdoctoral Researcher at the CHU Sainte Justine Research Center, University of Montreal, Canada. From 2024 to 2025, he was a Postdoctoral Research Associate at the Interdisciplinary Center for Security, Reliability, and Trust, University of Luxembourg, Luxembourg. He is currently a Postdoctoral Research Associate at the Conrad Blucher Institute, Texas A\&M University-Corpus Christi, USA. His research interests include applied AI/ML approaches for translationally and operationally critical decision-making systems. He was awarded Best Reviewer from the Machine Learning and Compression Workshop, NeurIPS 2024. He was a recipient of the merit FRQNT doctoral scholarship (2019-2023) awarded by the Fonds de recherche du Québec, Canada. He was also awarded a BrainKorea21$^+$ graduate scholarship by the Korean National Research Foundation, S. Korea.
\vspace{-5mm}
\end{IEEEbiography}

\begin{IEEEbiography}[{\includegraphics[width=0.8in, height=0.95in, clip, keepaspectratio]{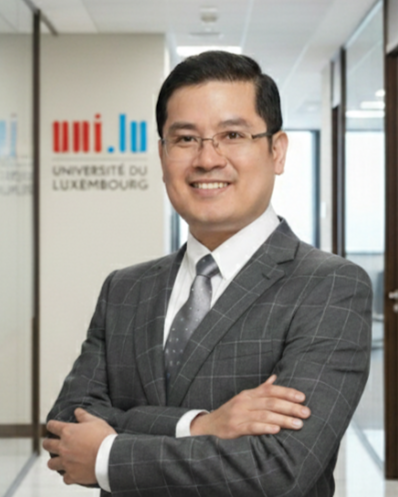}}]{Vu Nguyen Ha} (Senior Member, IEEE) received the B.Eng. degree (Hons.) from the French Training Program for Excellent Engineers in Vietnam, Ho Chi Minh City University of Technology, Vietnam, the Addendum degree from the École Nationale Supérieure des Télécommunications de Bretagne-Groupe des École des Télécommunications, Bretagne, France, in 2007, and the Ph.D. degree (Hons.) from the Institut National de la Recherche Scientifique-Énergie, Matériaux et Télécommunications, Université du Québec, Montreal, QC, Canada, in 2017. From 2016 to 2021, he worked as a Postdoctoral Fellow with the Ecole Polytechnique de Montreal, and then the Resilient Machine Learning Institute, École de Technologie Supérieure, University of Québec. He is currently a Research Scientist with the Interdisciplinary Centre for Security, Reliability, and Trust, University of Luxembourg. He was a recipient of the FRQNT Postdoctoral Fellowship for International Researcher (PBEEE) awarded by the Québec Ministry of Education, Canada, in 2018 and 2019. In 2021 and 2022, he was also awarded the Certificate for Exemplary Reviews by the IEEE Wireless Communications Letters.
\vspace{-5mm}
\end{IEEEbiography}

\begin{IEEEbiography}[{\includegraphics[width=0.8in, height=0.95in, clip, keepaspectratio]{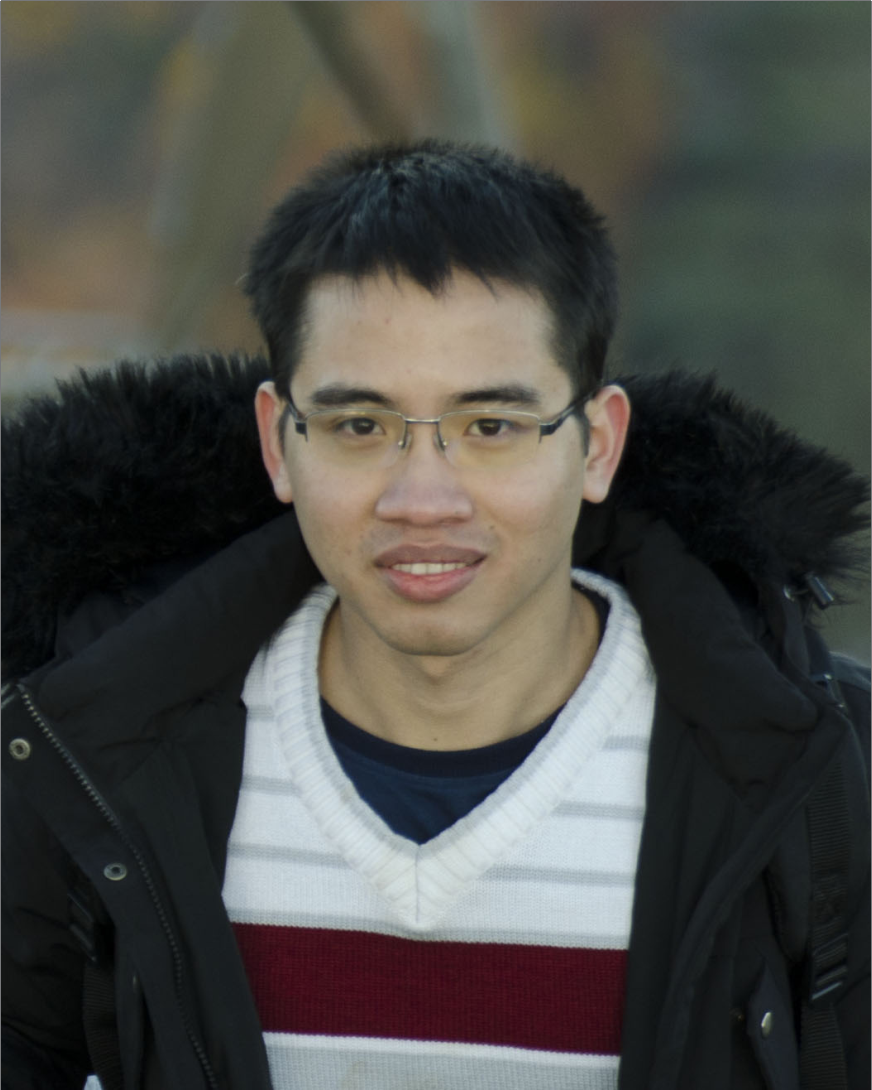}}]{Ti Ti Nguyen} (Member, IEEE) received the B.Eng. degree in electrical engineering from the Ho Chi Minh City University of Technology, Vietnam, in 2013, the M.Eng. degree in embedded system from the University of Rennes 1, France, in 2015, and the Ph.D. degree in telecommunications from Institut National de la Recherche Scientifique (INRS), Université du Québec, Canada, in 2020. From 2020 to 2024, he was a Postdoctoral Fellow with Synchromedia, École de Technologie Supérieure, Université du Québec, Canada. He is currently a Research Scientist with the Interdisciplinary Center for Security, Reliability and Trust (SnT), University of Luxembourg. His current research interests include SATCOM, edge computing, MIMO, NOMA, wideband communication, semantic communication, and AI/ML for wireless communications.
\vspace{-5mm}
\end{IEEEbiography}

\begin{IEEEbiography}[{\includegraphics[width=0.8in,height=0.95in,clip,keepaspectratio]{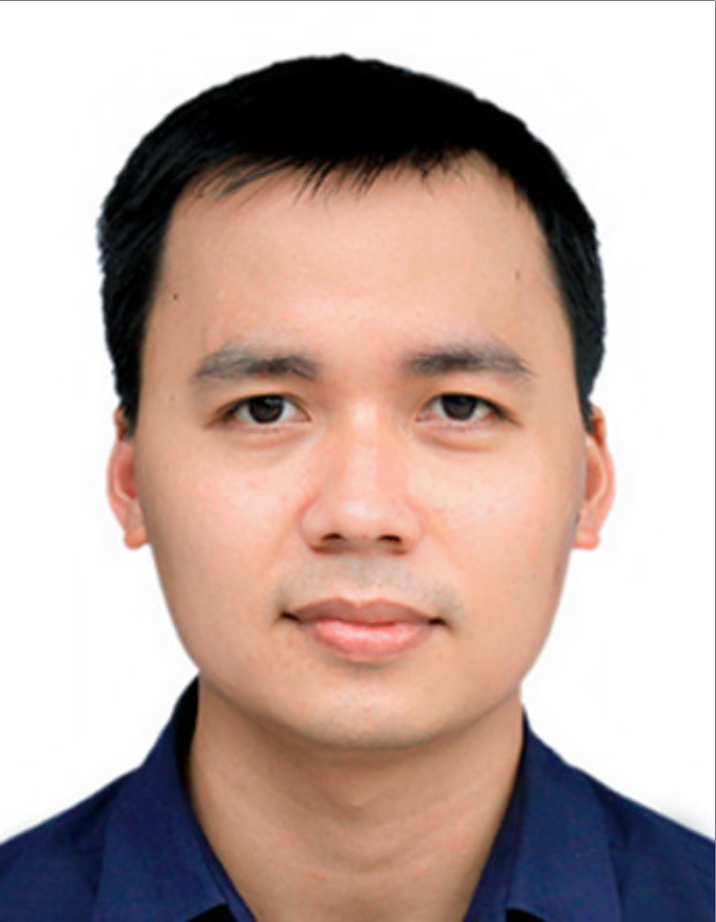}}]{Duc-Dung Tran} (Member, IEEE)  received the B.Eng. degree in electronics and telecommunications from Hue University of Sciences, Vietnam, in 2013, the M.Sc. degree in computer sciences from Duy Tan University, Vietnam, in 2016, and the Ph.D. degree in computer sciences from University of Luxembourg in 2024. From 2014 to 2019, he was with Faculty of Electrical and Electronics Engineering, Duy Tan University. He is currently a Research Associate at the Interdisciplinary Center for Security, Reliability and Trust (SnT), University of Luxembourg. His current research interests include 5G and beyond wireless networks, URLLC, multiple access techniques, and AI/ML for terrestrial and satellite communications.
\vspace{-5mm}
\end{IEEEbiography}

\begin{IEEEbiography}
	[{\includegraphics[width=0.8in,height=0.95in,clip,keepaspectratio]{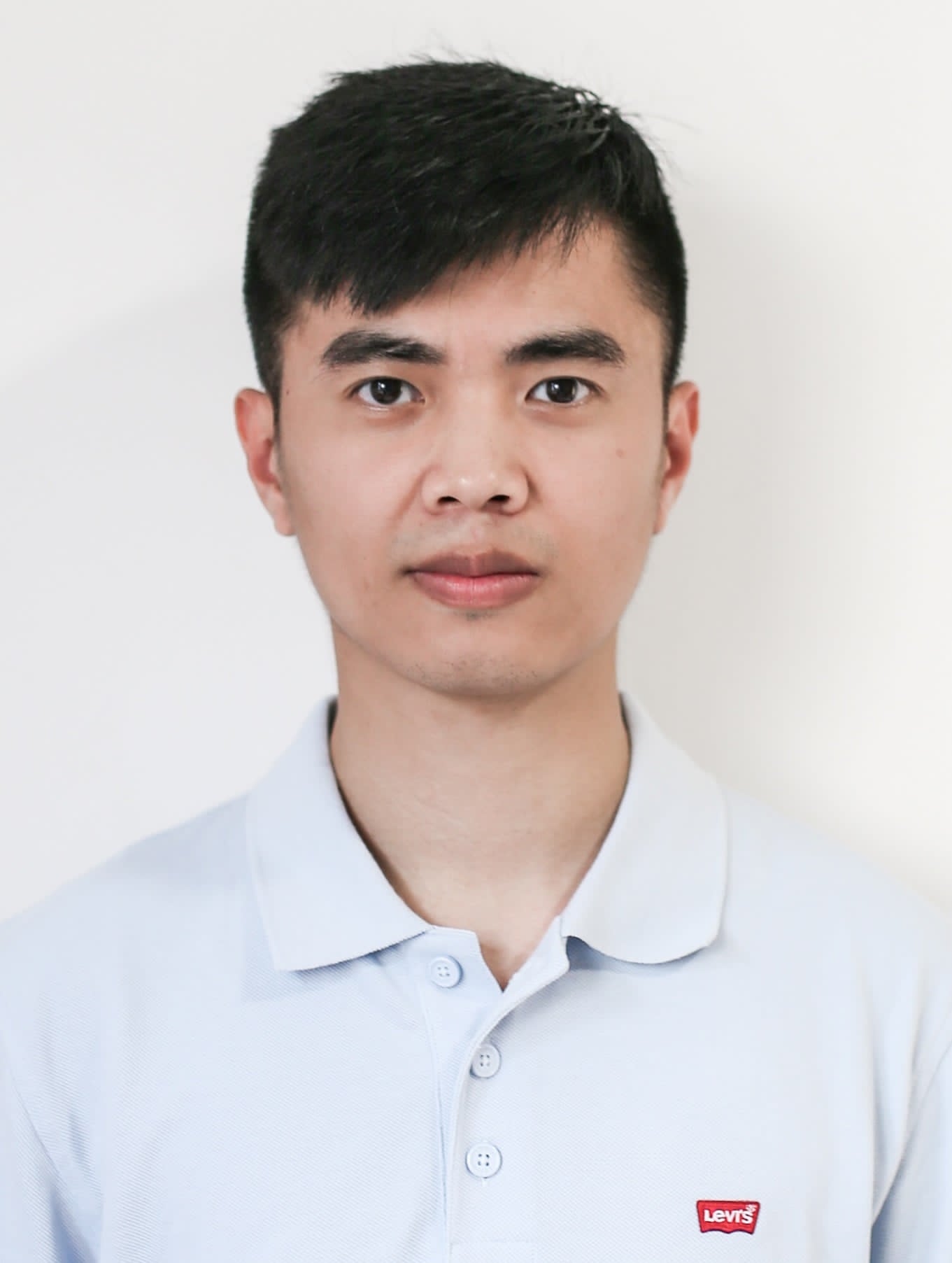}}]{Hung Nguyen-Kha} (Student Member, IEEE) received the B.E. degree in electronic and telecommunication from the Posts and Telecommunications Institute of Technology, Hanoi, Vietnam, in 2019, the M.S. degree in electrical engineering from Soongsil University, South Korea, in 2021, and Ph.D. degree in computer science from University of Luxembourg, Luxembourg, in 2025. He is currently working as a Research Associate with the Interdisciplinary Centre for Security, Reliability and Trust (SnT), University of Luxembourg. He received the Excellent Thesis Award for
	his Ph.D. degree in 2025, awarded by the University of Luxembourg.
	His research interests include applied optimization and machine-learning techniques for RRM problems in wireless communication systems, including SatCom, 5G/beyond-5G, massive MIMO.
\end{IEEEbiography}

\begin{IEEEbiography}[{\includegraphics[width=0.8in,height=0.95in,clip,keepaspectratio]{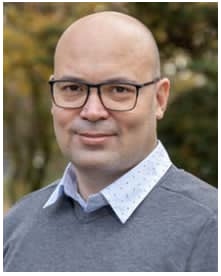}}]{Luis Manuel Garcés-Socarrás} (Member, IEEE) received the B.S. degree in automation control engineering, the M.S. degree in digital systems, and the Ph.D. degree in electronics from the Technological University of Havana, Cuba, in 2006, 2011, and 2017, respectively. From September 2008 to March 2022, he was a Lecturer and a Researcher with the Automation and Computing Department, Technological University of Havana. He was a Visiting Researcher at the Seville Institute of Microelectronics (IMSE-CNM), Spain, in 2010, 2011, and 2013. From September 2014 to March 2015, he was a Visiting Researcher with the Microelectronics Group at the Federal University of Itajubá (UNIFEI), Brazil, and from September to December 2017, he held a Postdoctoral Researcher position at the same institution. Since April 2022, he has been with the Interdisciplinary Centre for Security, Reliability, and Trust (SnT), University of Luxembourg, where he worked as a Research and Development Specialist and, since January 2024, as a Research Associate/Postdoctoral Researcher. His research interests include digital signal processing, FPGA-based implementations, and embedded systems for satellite communication applications. \end{IEEEbiography}

\begin{IEEEbiography}[{\includegraphics[width=0.8in,height=0.95in,clip,keepaspectratio]{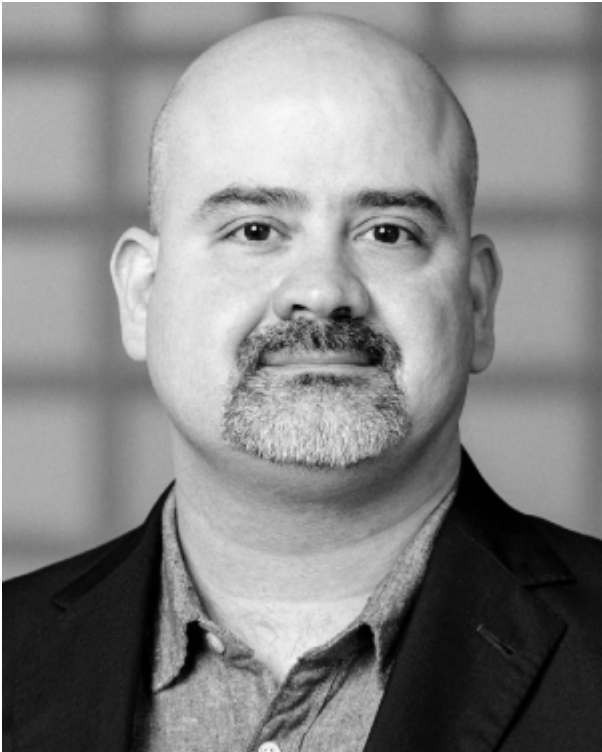}}]{Juan Carlos Merlano-Duncan} (Senior Member, IEEE) received the Diploma degree in electrical engineering from Universidad del Norte, Barranquilla, Colombia, in 2004, and the M.Sc. and Ph.D. Diploma (cum laude) degrees from Universitat Politecnica de Catalunya (UPC), Barcelona, Spain, in 2009 and 2012, respectively. At UPC, he was responsible for the design and implementation of a radar system known as SABRINA, the first ground-based bistatic radar receiver using space-borne platforms such as ERS-2, ENVISAT, and TerraSAR-X as opportunity transmitters (C and X bands). He implemented a ground-based array of transmitters, which monitored land subsidence with subwavelength precision. These two implementations involved FPGA design, embedded programming, and analog RF/Microwave design. In 2013, he joined the Institut National de la Recherche Scientifique, Montreal, Canada, as a Research Assistant to design and implement cognitive radio networks using software development and FPGA programming. He has been with the SIGCOM research group of the Interdisciplinary Center for Security, Reliability, and Trust (SnT), University of Luxembourg, Luxembourg, since 2016, where he currently works as a Research Scientist leading the COMMLAB laboratory at SnT working on SDR implementation of satellite and terrestrial communication systems. His research interests include wireless communications, remote sensing, distributed systems, frequency distribution and carrier synchronization systems, software-defined radios, and embedded systems.
\end{IEEEbiography}

\begin{IEEEbiography}[{\includegraphics[width=0.8in, height=0.95in, clip, keepaspectratio]{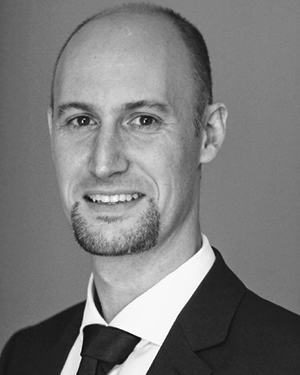}}]{Symeon Chatzinotas} (Fellow, IEEE) received the M.Eng. degree in telecommunications from the Aristotle University of Thessaloniki, Greece, in 2003, and the M.Sc. and Ph.D. degrees in electronic engineering from the University of Surrey, U.K., in 2006 and 2009, respectively. He is currently a Full Professor/Chief Scientist I and the Head of the Research Group SIGCOM, Interdisciplinary Centre for Security, Reliability and Trust, University of Luxembourg. In parallel, he is an Adjunct Professor with the Department of Electronic Systems, Norwegian University of Science and Technology and a Collaborating Scholar with the Institute of Informatics and Telecommunications, National Center for Scientific Research “Demokritos.” In the past, he has lectured as a Visiting Professor with the University of Parma, Italy and contributed in numerous research and development projects for the Institute of Telematics and Informatics, Center of Research and Technology Hellas and the Mobile Communications Research Group, Center of Communication Systems Research, University of Surrey. He has authored more than 700 technical papers in refereed international journals, conferences, and scientific books and has received numerous awards and recognitions, including the IEEE Fellowship and an IEEE Distinguished Contributions Award. He is currently on the editorial board of the IEEE Transactions on Communications, IEEE Open Journal of Vehicular Technology, and the International Journal of Satellite Communications and Networking.
\vspace{-5mm}
\end{IEEEbiography}

\end{document}